%% file: main.tex
\documentclass[9pt,twocolumn]{extarticle}

\usepackage{graphicx,amsmath,amssymb}
\usepackage[super,comma,sort&compress]{natbib}
\setcitestyle{nature}
\usepackage[breaklinks=true,colorlinks,allcolors=blue,bookmarks]{hyperref}

\usepackage{macros/my_macros}
\usepackage{macros/my_formatting}
\usepackage{macros/popupcite}

\input{main.bbl}
\input{supplementary.bbl}
\begin{document}

\input{cap.tex}
\input{introduction.tex}
\input{results.tex}
\input{discussion.tex}
\input{methods.tex}
\input{trivia.tex}

\input{main.bbl}
\input{supplementary.tex}
\fontsize{8.9}{10.9}\selectfont

\input{supplementary.bbl}
\end{document}

%% file: cap.tex
\title{Tunnel field-effect transistors for sensitive terahertz detection}

\author[1,2][1]{I.~Gayduchenko}
\author[3,4][1]{S.G.~Xu}
\author[1][1]{G.~Alymov}
\author[1,2]{M.~Moskotin}
\author[5]{I.~Tretyakov}
\author[6]{T.~Taniguchi}
\author[7]{K.~Watanabe}
\author[2,8]{G.~Goltsman}
\author[3,4]{A.K.~Geim}
\author[1,2][2]{G.~Fedorov}
\author[1][2]{D.~Svintsov}
\author[1,3,9][2]{D.A.~Bandurin}

\affil[1]{Moscow Institute of Physics and Technology (National Research University), Dolgoprudny 141700, Russia}
\affil[2]{Physics Department, Moscow Pedagogical State University, Moscow, 119435, Russia}
\affil[3]{School of Physics, University of Manchester, Oxford Road, Manchester M13 9PL, United Kingdom}
\affil[4]{National Graphene Institute, University of Manchester, Manchester M13 9PL, United Kingdom}
\affil[5]{Astro Space Center, Lebedev Physical Institute of the Russian Academy of Sciences, Moscow 117997, Russia}
\affil[6]{International Center for Materials Nanoarchitectonics, National Institute of Material Science, Tsukuba 305-0044, Japan}
\affil[7]{Research Center for Functional Materials, National Institute of Material Science, Tsukuba 305-0044, Japan}
\affil[8]{National Research University Higher School of Economics, Moscow, 101000, Russia}
\affil[9]{Present address: Department of Physics, Massachusetts Institute of Technology, Cambridge, Massachusetts 02139, USA}

\thanks[1]{Equally contributed authors.}
\thanks[2]{Corresponding authors: \mailto{gefedorov@mail.ru}, \mailto{svintcov.da@mipt.ru}, \mailto{bandurin@mit.edu}.}

\abstract{
The rectification of electromagnetic waves to direct currents is a crucial process for energy harvesting, beyond-5G wireless communications, ultra-fast science, and observational astronomy. As the radiation frequency is raised to the sub-terahertz (THz) domain, ac-to-dc conversion by conventional electronics becomes challenging and requires alternative rectification protocols. Here we address this challenge by tunnel field-effect transistors made of bilayer graphene (BLG). Taking advantage of BLG's electrically tunable band structure, we create a lateral tunnel junction and couple it to an antenna exposed to THz radiation. The incoming radiation is then down-converted by the tunnel junction nonlinearity, resulting in high-responsivity ($>4$~kV/W) and low-noise (0.2 pW/$\sqrt{\mathrm{Hz}}$) detection. We demonstrate how switching from intraband Ohmic to interband tunneling regime can raise detectors' responsivity by few orders of magnitude, in agreement with the developed theory. Our work demonstrates a potential application of tunnel transistors for THz detection and reveals BLG as a promising platform therefor.
}

\maketitle

%% file: introduction.tex
\begin{introduction}[F]ield effect transistors (FETs) recently found an unexpected application for the  rectification of THz and sub-THz signals beyond their cutoff frequency~\cite{DS1996,knap2009field}. This technology paves the way for on-chip~\cite{Lisauskas_CMOS_detectors}, low-noise~\cite{GaN_low_noise}, and sub-nanosecond radiation detection~\cite{viti2020thermoelectric,muravev2016response} offering the possibility of $\gtrsim 10$ Gb/s data transfer rates. Contrary to competing diode rectifiers, FETs offer the possibility of phase-sensitive detection~\cite{Phase_sensitive,Heterodyne_FET} vital for noise-immune communications with phase modulated signals. Recent innovations towards enhanced responsivity of FET-detectors include the use of 2D materials~\cite{VitielloNanowires,Vicarelli2012,BPTHz}, exotic nonlinearities~\cite{FUTHZ,RatchetGanichev,KC-JJ-microwave,GregTHz,pseudo-Euler}, enhanced light-matter coupling~\cite{ThermoTHzKoppnesNanoL} and plasmonic effects~\cite{MuravevDefect,Bandurin2018,CorbinoTHz}. Despite the rich and complex physics of THz rectification, the responsivity of most FET-detectors is governed by the sensitivity of the channel conductivity $G_\mathrm{ch}$ to the gate voltage $V_\mathrm{g}$, parameterized via the normalized transconductance $F=-d \ln G_\mathrm{ch}/d V_\mathrm{g}$~\cite{knap2009field,Sakowicz}. The transconductance in conventional FETs has a fundamental limit of $e/k_B T$ ($e$ is the elementary charge and $k_B T$ is the thermal energy) dictated by the leakage of thermal carriers over the gate-induced barrier, termed as 'Boltzmann tyranny'. Although this process is well-recognized as a limiting factor for the minimal power dissipation of FETs in integrated circuits, it has been scarcely realised that it also imposes a bound on the responsivity of antenna-coupled FETs to THz fields.  

One of the most promising routes to escape from the Boltzmann tyranny is the manipulation of interband tunneling instead of intraband thermionic currents. This idea is materialized in a tunnel field effect transistor (TFET)~\cite{Ionescu,Sarkar,Seabaugh_state_of_the_art}. TFETs find their applications in low-voltage electrical and optical switching~\cite{Tunneling_switch}, accelerometry~\cite{Tunneling_accelerometer}, chemical~\cite{TFET_gas_sensor} and biological sensing~\cite{Biosensor_TFET,TFET_biosensor_experiment}.
In spite of this variety, the use of TFETs for the rectification of high-frequency signals~\cite{TFET-THz-IEEE} has not been attempted so far. This is also surprising considering recent advances in the development of tunnelling high-frequency rectifiers and detectors based on quantum dots~\cite{QD-THz1,QD-THz2}, diodes~\cite{TunnelingIII-V-THz,THz-Tunnel_rings,THz-receiversSciRep,Rectenna1,rectenna2,rectenna3} and superconducting tunnel junctions~\cite{SC-TJ1,SC-TJ-graphene,KCFONG}. A possible reason is that low on-state current and cut-off frequency of TFETs stimulate the belief on their inapplicability in teratronics~\cite{Analog_performance_TFET}.

In this work, we show that the opposite is true and demonstrate the use of TFETs for highly-sensitive sub-THz and THz detection. Using bilayer graphene (BLG) as a convenient platform for this enquiry, we fabricate a dual-gated TFET and couple it to a broadband THz antenna. The received high-frequency signal is rectified by electrostatically-defined tunnel junction resulting in high-responsivity and low-noise detection. Our experimental results and the developed theory suggest that the origin of the high responsivity in our detectors is not the large transcoductance, but rather steep curvature of the tunneling $I-V$ characteristic~\cite{Ryzhii_lateral_Schottky}. Our findings point out that even TFETs without sub - $k_B T/e$ switching can act as efficient THz rectifiers preserving all the benefits of transistor-based detection technology.
\end{introduction}

%% file: results.tex
\section{Results}
\subsection{Device fabrication and characterization.}
\begin{figure*}[ht]
	\centering\includegraphics[width=0.97\linewidth]{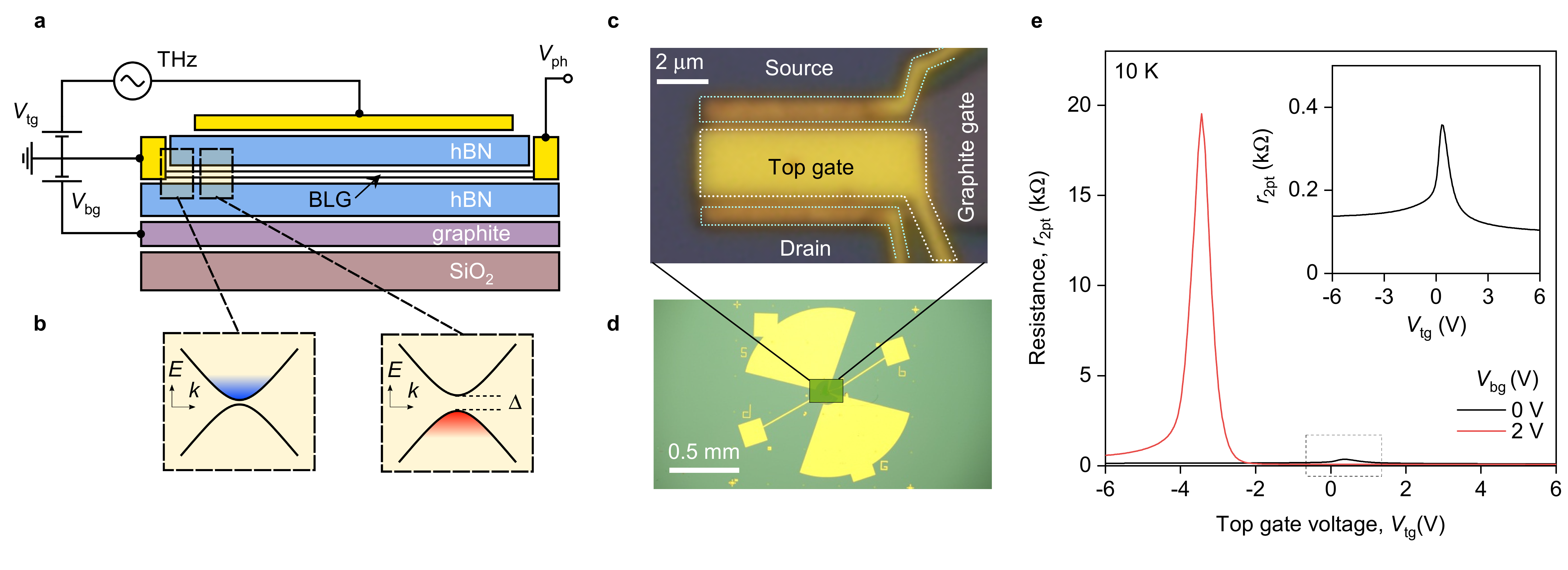}
	\caption{\textbf{Dual-gated bilayer graphene THz detector. }\textbf{a,} Schematic of an hBN encapsulated dual-gated BLG transistor. THz radiation is incident on a broadband antenna connected to the source (S) and gate terminals yielding modulation of the top gate-to-source voltage ($V_\mathrm{tg}$) while the back gate voltage ($V_\mathrm{bg}$) is fixed. The build-up photovoltage $V_\mathrm{ph}$ is read out between the source and drain (D) terminals.  \textbf{ b,} Band structure of the BLG at the interface between the $n$-doped bottom gate-sensitive region and dual-gated $p$-doped channel ($\Delta$ is the band gap). Blue and red colours illustrate conduction and valence bands fillings respectively. Note, even for a single-gated region, a finite band gap appears in the energy dispersion due to the difference in on-site energies between the top and bottom graphene layers~\cite{VolodyaPRL-BLG}. \textbf{c-d,} Optical photographs of the fabricated dual-gated detector.  The source and top-gate terminals are connected to a broadband bow-tie antenna.  \textbf{e,} The two-terminal resistance of our BLG device, $r_\mathrm{2pt}$, as a function of $V_\mathrm{tg}$ for two representative $V_\mathrm{bg}=0$ and $V_\mathrm{bg}=2$~V. Inset: Zoomed-in $r_\mathrm{2pt}(V_\mathrm{tg})$ for $V_\mathrm{bg}=0$~V. $T=10$ K.
}
\label{Fig1}
\end{figure*}
\begin{figure*}[ht!]
	\centering\includegraphics[width=0.79\linewidth]{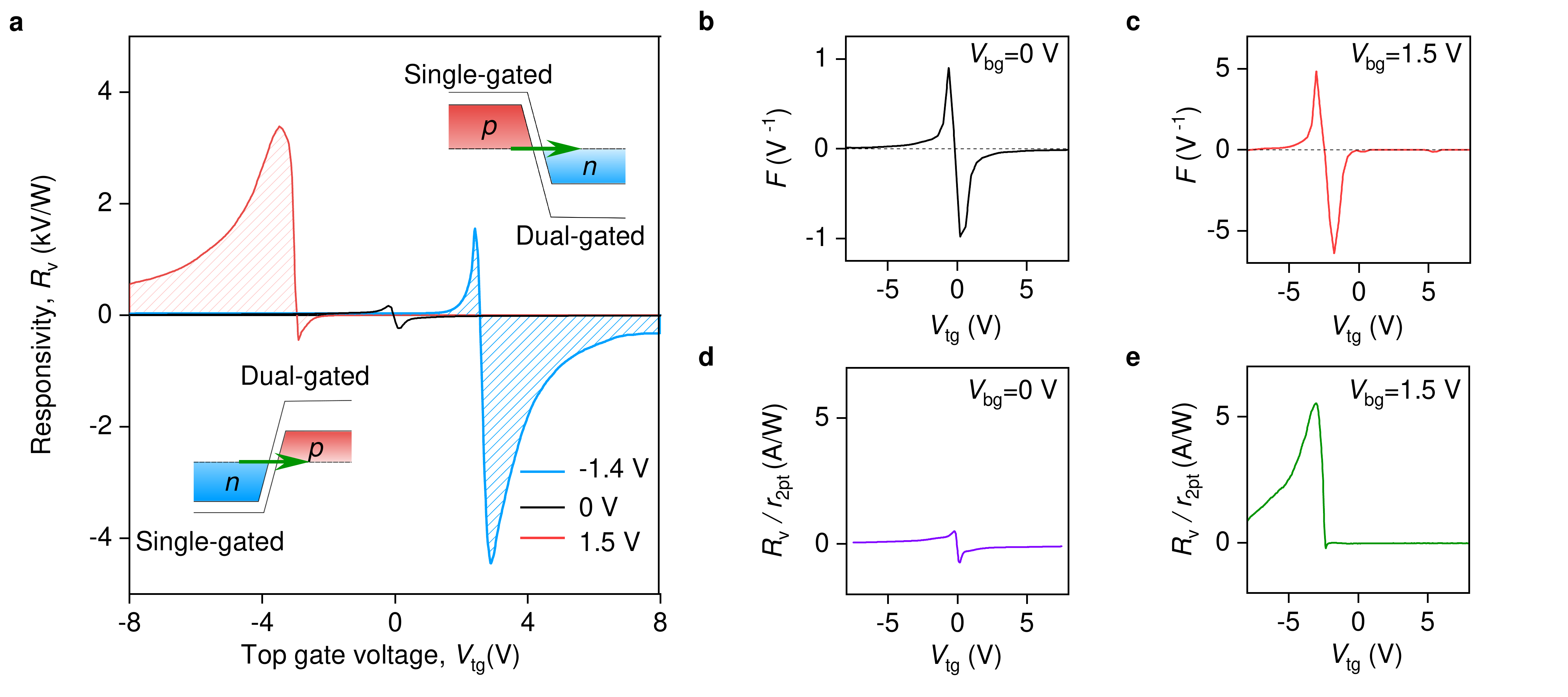}
	\caption{\textbf{Tunnelling-assisted THz detection} \textbf{a,} Detector responsivity, $R_\mathrm{v}$, as a function of $V_\mathrm{tg}$ for $V_\mathrm{bg}=0$~V (black), $V_\mathrm{bg}=-1.4$~V (blue) and $V_\mathrm{bg}=1.5$~V (red) measured in response to $f=0.13$~THz radiation. $T=10$~K. Inset illustrates band profiles in the vicinity of the single and dual-gated interface when $V_\mathrm{bg}$ and $V_\mathrm{tg}$ are of opposite polarities. Green arrow illustrates interband tunnelling.  \textbf{b,c} Normalized transconductance $F$ versus $V_\mathrm{tg}$ obtained by numerical differentiation of the device resistance for $V_\mathrm{bg}=0~$V (b) and $V_\mathrm{bg}=1.5~$V (c). Note, $F(V_\mathrm{tg})$ dependencies are fairly symmetric whereas the $R_\mathrm{v}(V_\mathrm{tg})$ is highly asymmetric for the same $V_\mathrm{bg}$ (a). \textbf{d, e} $R_\mathrm{v}$ from (a) normalized to the channel resistance $r_\mathrm{2pt}$ as a function of $V_\mathrm{tg}$ for given $V_\mathrm{bg}$.
	}
	\label{Fig2}
\end{figure*}

\begin{figure*}[ht!]
	\centering\includegraphics[width=0.84\linewidth]{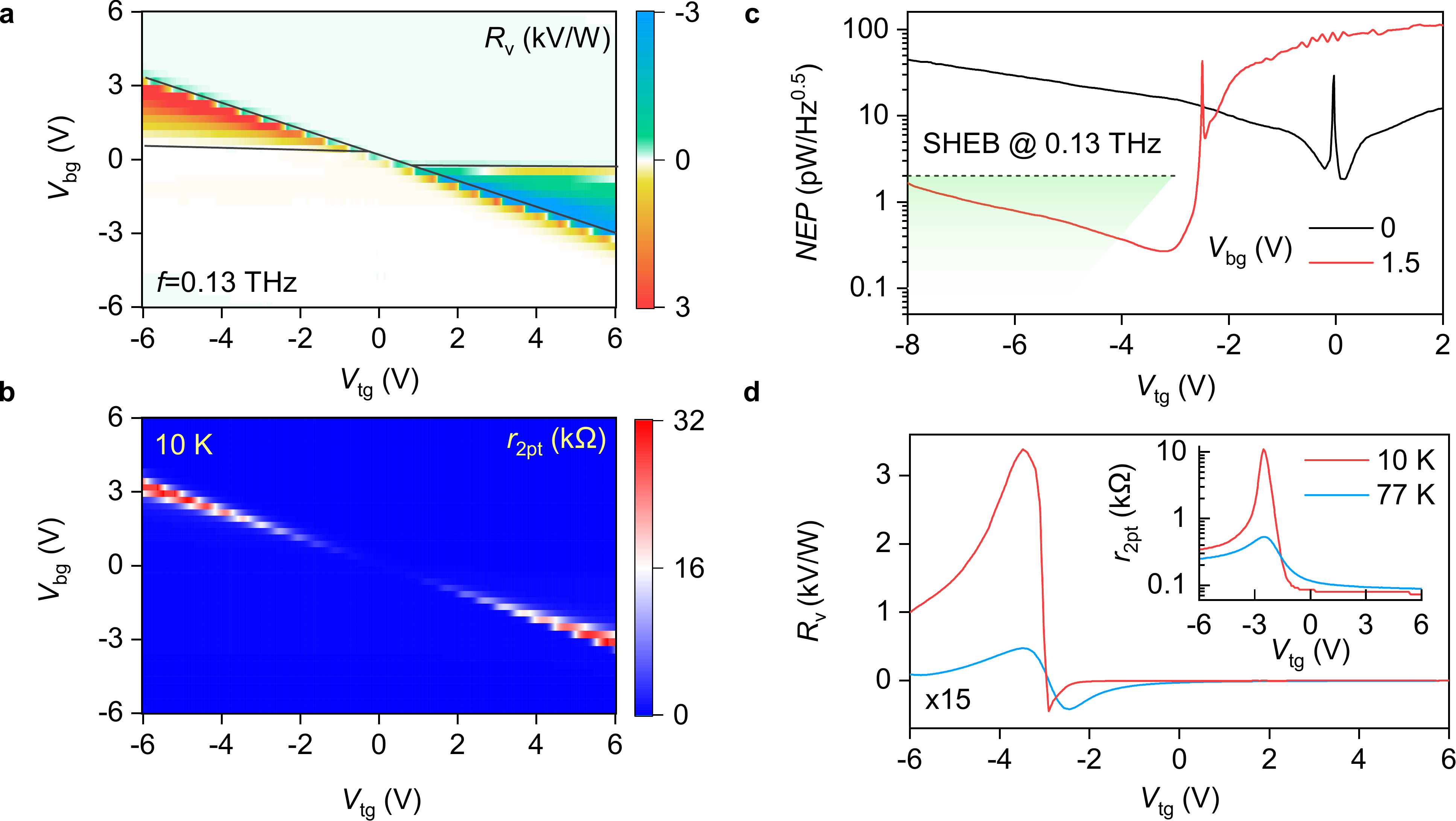}
	\caption{\textbf{Performance of the BLG-TFET detector.} \textbf{a,} Responsivity of our detector as a function of $V_\mathrm{bg}$ and $V_\mathrm{tg}$ recorded in response to $f=0.13$ THz radiation. The black lines demark the $(V_\mathrm{tg},V_\mathrm{tg})$ regions where the tunnel junction configuration is realized. \textbf{b,} $r_\mathrm{2pt}(V_\mathrm{tg},V_\mathrm{tg})$ map measured at $T=10$ K. The appearance of highly resistive regions (red) points out to the band gap opening in BLG. \textbf{c,} NEP of our detector at given $V_\mathrm{bg}$ determined using the Johnson--Nyquist relation for the noise spectral density. Horizontal line marks NEP level for SHEBs operating at the same $f$ and $T=4.2$~K (See \ref{Section:Comparison} for a detailed comparison of the BLG-TFET with other THz detectors). Green shaded region indicates the spread in NEP for SHEBs at higher $f$. \textbf{d,} Temperature dependence of $R_\mathrm{v}(V_\mathrm{tg})$ and $r_\mathrm{2pt}(V_\mathrm{tg})$ (inset) at $V_\mathrm{bg}=1.5$~V. 
	}
	\label{Fig3}
\end{figure*}

For the proof-of-principle demonstration, we constructed a TFET of a BLG taking advantage of its unique electronic properties. BLG is a narrow-band semiconductor characterized by a tunable band structure highly sensitive to the transverse electric field~\cite{VolodyaPRL-BLG}. This ensures a steep ambipolar field effect and allows for an independent control of the band gap size and the carrier density, $n$~\cite{BLG-Yanbo}, providing a unique opportunity for a fully electrostatic engineering of the spatial band profile~\cite{Banszerus2018,EnsslinQD2,EnsslinQD}. We employed this property to electrostatically define typical TFET configuration shown in Fig.~\ref{Fig1}a,b. In addition, BLG hosts a high-mobility electronic system, a crucial property for high-frequency applications. As we now proceed to show, these properties make BLG a convenient platform to demonstrate the drastic differences in performance of intraband field-effect-enabled detection and its interband tunneling counterpart within the same device. 

We fabricated our detector by an encapsulation of BLG between two slabs of hexagonal boron nitride (hBN) using standard dry transfer technique described elsewhere~\cite{Kretinin} (See \nameref{sec:methods}). The BLG channel of length $L$ = 2.8 $\mu$m and width $W$ = 6.2 $\mu$m was assembled on top of a relatively thin ($\sim10$ nm) graphite back gate which ensured efficient screening of remote charge impurities in Si/SiO$_2$ substrate~\cite{ZibrovNature}. The device was also equipped with a second (top) gate electrode deposited symmetrically between the source and drain contacts. Importantly, relatively short ($l<100$ nm) regions near the contacts were not covered by the top gate and thus were affected by the bottom one only. This configuration allowed us to define a lateral tunnel junction between single- and double-gated regions when the top and bottom gate voltages ($V_{\mathrm{tg}}$ and $V_{\mathrm{bg}}$ respectively) had opposite polarities~\cite{Banszerus2018,EnsslinQD2,EnsslinQD}, as explained in Figs.~\ref{Fig1}b and \ref{FigECircuit}. The device was coupled to the incident radiation via broadband bow-tie antenna connected to the source and top gate electrodes. The rectified dc photovoltage, $V_\mathrm{ph}$, was read out between the source and drain terminals as shown in Fig.~\ref{Fig1}a (See \nameref{sec:methods}).

Prior to photoresponse measurements, we characterized the transport properties of our device. Figure~\ref{Fig1}e shows the dependence of our detector's two-terminal resistance, $r_\mathrm{2pt}$, on $V_\mathrm{tg}$ for two representative values of $V_\mathrm{bg}$ measured at $T=10$~K. At $V_\mathrm{bg}=0$~V, $r_\mathrm{2pt}(V_\mathrm{tg})$ exhibits familiar bell-like structure that peaks at the charge neutrality point (CNP) where $r_\mathrm{2pt}\approx0.4$ k$\Omega$ (inset of Fig.~\ref{Fig1}e). Application of $V_\mathrm{bg} = 2$ V shifts the CNP to negative $V_\mathrm{tg}$ and results in drastic increase of $r_\mathrm{2pt}$ that reaches $20~\mathrm{k}\Omega$ already at $V_\mathrm{tg}\approx-3.5$~V. This increase is a clear indicative of the electrically-induced band gap in BLG~\cite{VolodyaPRL-BLG,BLG-Yanbo}.

\subsection{Tunneling-enabled detection.}
\begin{figure*}[ht!]
	\centering\includegraphics[width=0.85\linewidth]{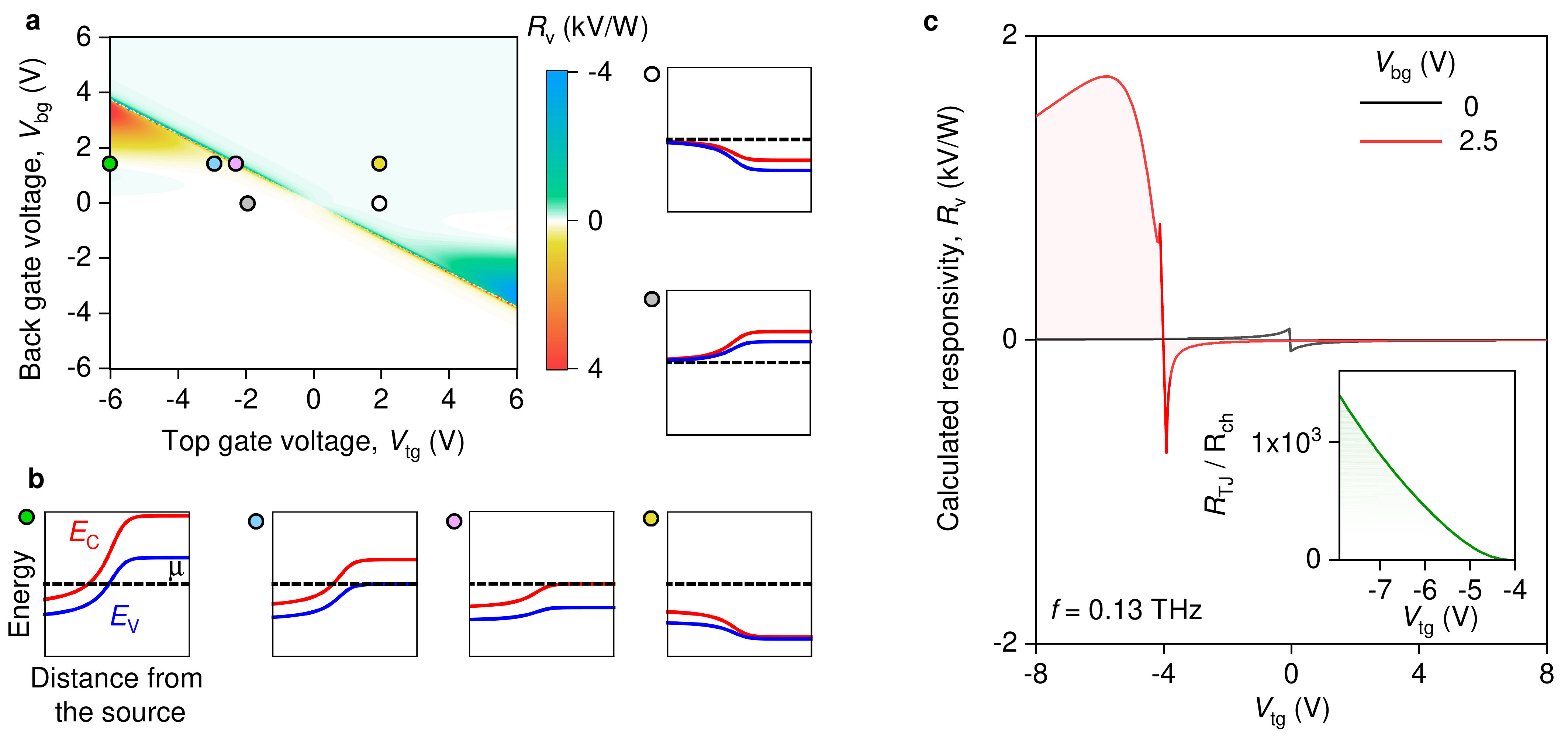}
	\caption{\textbf{Modelling tunneling-assisted THz detection.} \textbf{a,} Calculated $R_\mathrm{v}(V_\mathrm{tg},V_\mathrm{bg})$ map of our dual-gated BLG device in response to $f=0.13~$THz radiation. \textbf{b,} Calculated band profiles for different $(V_\mathrm{tg},V_\mathrm{bg})$ configurations indicated by the colored symbols in (a). White, grey, yellow and pink symbols point to the band diagrams of the FET mode whereas the green and blue symbols correspond to the regime of interband tunnelling. Red, blue and black lines illustrate conduction band minimum ($E_\mathrm{C}$), valence band maximum ($E_\mathrm{V}$), and the chemical potential ($\mu$), respectively. \textbf{c,} Line cuts of the map in (a) for $V_\mathrm{bg}=0$~V and $V_\mathrm{bg}=2.5$~V. The radiation resistance of the antenna $Z_{\rm rad}\approx75~\Omega$ was used for these calculations (\nameref{sec:methods}). Inset: The ratio between the tunnel junction, $R_\mathrm{TJ}$, contribution to the responsivity  and that of the channel nonlinearity, $R_\mathrm{ch}$, for $V_\mathrm{bg}=2.5$~V. }
	\label{Fig4}
\end{figure*}
Figure~\ref{Fig2}a shows the external responsivity of our detector, $R_\mathrm v=V_\mathrm{ph}/P_{\rm in}$, as a function of $V_\mathrm{tg}$ recorded in response to $f=0.13$~THz radiation. Here $V_\mathrm{ph}$ is the generated photovoltage and $P_{\rm in}$ is the incident radiation power (See \nameref{sec:methods} for the details of responsivity determination).  At $V_\mathrm{bg}=0$~V, $R_\mathrm v(V_\mathrm{tg})$ exhibits a standard antisymmetric sign-changing behaviour with $|R_\mathrm v|$ reaching 200~V/W close to the CNP. The functional form of $R_\mathrm v(V_\mathrm{tg})$ follows that of the normalized transconductance $F=-(d G_\mathrm{ch}/d V_\mathrm{tg})/G_\mathrm{ch}$ (Fig.~\ref{Fig2}b), where $G_\mathrm{ch}=1/r_\mathrm{2pt}$, and is consistent with previous studies of graphene-based THz detectors~\cite{Vicarelli2012,SpiritoBLG,Bandurin2018}. This standard behaviour is routinely understood in terms of a combination of resistive self-mixing and photo-thermoelectric rectification, two predominant mechanisms that govern THz detection in graphene-based FETs~\cite{BandurinAPL-THz}.  

The response of our device changes drastically when a finite vertical electric field is applied perpendicular to the BLG channel. Figure~\ref{Fig2}a shows the $R_\mathrm v (V_\mathrm {tg})$ dependence for $V_\mathrm {bg}=1.5~V$ and reveals a giant increase of $R_\mathrm v$ exceeding 3 kV/W (red curve). A notable feature of the observed dependence is its strong asymmetry with respect to zero $V_\mathrm{bg}$ behaviour: namely, $|R_\mathrm v|$ is more than an order of magnitude larger for the $p$-doped channel (to the left from the CNP in Fig.~\ref{Fig1}e) as compared to the case of $n$-doping (to the right from the CNP in Fig.~\ref{Fig1}e). In addition, while the response decays rapidly with increasing $V_\mathrm{tg}$ on the $n$-doped side, a finite $R_\mathrm v$ is observed over the whole span of $V_\mathrm {tg}$ at which the channel is $p$-doped. Furthermore, when the sign of $V_\mathrm{bg}$ is reversed, $R_\mathrm{v}(V_\mathrm{tg})$ remains asymmetric but, in this case, it is strongly enhanced for positive $V_\mathrm{tg}$ (blue curve in Fig.~\ref{Fig2}a). Importantly, the $F$-factor remains fairly symmetric for the $V_\mathrm {tg}/V_\mathrm {bg} $ combinations at which $R_\mathrm v$ exhibits strong asymmetry. This observation suggests that the strong rectification of THz radiation in our device is not caused by the non-linearities in the dual-gated BLG channel. Moreover, the increase in $R_\mathrm v$ cannot be explained by a trivial enhancement of channel resistance.
To demonstrate this, in Fig.~\ref{Fig2}d and e we plot $R_{\rm v}$ normalized to $r_{\rm 2pt}$, a quantity with the dimension of current responsivity. A symmetric $R_{\rm v} / r_{\rm 2pt}(V_{\rm tg})$ dependence measured at $V_{\rm bg}=0$~V is conceded with amplified and highly-asymmetric curve at finite $V_{\rm bg}$, thereby excluding resistance-enabled $R_{\rm v}$ enhancement.

Figure~\ref{Fig3}a-b details our observations further by showing maps of $R_\mathrm v (V_\mathrm {tg},V_\mathrm {bg})$ and $r_\mathrm {2pt} (V_\mathrm {tg},V_\mathrm {bg})$. Enhanced $R_\mathrm {v}$ is observed in two distinct quadrants characterized by an anti-symmetric (with respect to the $V_\mathrm {bg}$) sign pattern (see \ref{Section:Further_examples} for the line cuts of the map in Fig.~\ref{Fig3}a). Outside these quadrants, $R_\mathrm v$ was found negligibly small. Interestingly, $r_\mathrm {2pt}$-map is fairly symmetric featuring gradual increase of resistance at the CNP with increasing vertical field as expected for  BLG~\cite{VolodyaPRL-BLG,BLG-Yanbo}. We have also studied the performance of our detectors at higher $f$ and found consistent highly asymmetric response similar to that shown in Fig.~\ref{Fig2}b (\ref{Section:Broadband_photoresponse}) highlighting broadband character of the rectification mechanism. Furthermore, using Johnson--Nyquist relation for the noise spectral density $S=\sqrt{4k_\mathrm{B}Tr_\mathrm{2pt}}$, we estimate the noise equivalent power of our detector, NEP=$S/R_\mathrm v$, to reach 0.2 pW$/\sqrt{\mathrm{Hz}}$ at $T \approx 10$~K (Fig.~\ref{Fig3}c). For comparison, commercial superconducting hot electron bolometers (SHEB) operating at lower $T=4.2~$K feature NEP of 0.1-2 pW/$\sqrt{\mathrm{Hz}}$ (Fig.~\ref{Fig3}c, green shaded area) that makes our dual-gated detectors competitive with the commercial technology. (\ref{Section:Comparison}).

In order to understand deeper the peculiar detection mechanism of our cooled detector, we have studied the temperature dependence of its performance. Figure~\ref{Fig3}d compares the $R_\mathrm {v}(V_\mathrm {tg})$ dependencies measured at $T=10$ and 77 K in response to $f=0.13$ THz radiation. For $p$-doped channel, $R_\mathrm {v}(V_\mathrm {tg})$ drops by more than 2 orders of magnitude whereas a 10-fold decrease in $R_\mathrm {v}(V_\mathrm {tg})$ is observed for the $n$-doped side. Furthermore, in  contrast to the behaviour observed at $10$ K, $R_\mathrm {v}(V_\mathrm {tg})$ curves become more symmetric at liquid nitrogen $T$. To compare, $r_\mathrm {2pt}$ at the CNP also drops with increasing $T$ (inset of Fig.~\ref{Fig3}d), demonstrating usual insulating behaviour of gapped BLG at zero doping. However, $r_\mathrm {2pt}$ exhibits clear asymmetry with respect to the CNP. In particular, for the case of $n$-doped channel, $r_\mathrm {2pt}$ is rather small ($\approx100~\Omega$) and it grows with increasing $T$, a signature of phonon-limited transport, whereas on the $p$-doped side we observed a pronounced decrease of $r_\mathrm {2pt}$ with increasing $T$; $r_\mathrm {2pt}$ is of the order of 0.5~k$\Omega$ away from the CNP. The giant enhancement of $R_\mathrm {v}$, insulating $T$-dependence of $r_\mathrm {2pt}$ and its increase, when $V_\mathrm {tg}$ and $V_\mathrm {bg}$ are of opposite polarities, suggest that the behaviour of our BLG detector is governed by the interband tunneling as we now proceed to demonstrate.

\subsection{Modelling tunneling-enabled photoresponse.}
Our dual-gated BLG transistor can be modelled by an equivalent circuit described in \nameref{sec:methods} (see below).
It consists of the gate-controlled channel conductance $\GCSG$ and tunnel junctions at the source and drain with conductances $G_{\rm S}$ and  $G_{\rm D}$, respectively. The net responsivity $R_\mathrm{v}$ of such a circuit is the sum of three 'intrinsic' responsivities (marked with subscript $_{i}$) weighted with voltage division factors $\gamma = [1+ G_{\rm S}/\GCSG]^{-1}$ and multiplied by a factor of $\approx 4Z_\mathrm{rad}$ relating the mean square of the antenna's output voltage with the incident power ($Z_\mathrm{rad}$ is the radiation resistance of the antenna; exact expression for the prefactor is given in \nameref{sec:methods}):
\begin{equation}
\label{Responsivity_main}
    R_\mathrm{v} \approx 4Z_\mathrm{rad}[\RSS \abs{\gamma}^2 + \RSG \Re\gamma + \RC \abs{1-\gamma}^2].
\end{equation}
The channel responsivity, $\RC$, is proportional to the transconductance~\cite{Sakowicz} and appears due to resistive self-mixing effect, i.e. due to simultaneous modulation of carrier density by transverse gate field and their drag by longitudinal field. The tunnel junction responsivity $\RSS$ emerges due to non-linear dependence of tunneling current on junction voltage, $V_{\rm TJ}$. Finally, the responsivity $\RSG$ appears due to the simultaneous action of the gate voltage that modulates tunnel barrier and junction voltage that pulls the carriers. All three contributions can be calculated from the sensitivities of conductances $G_{\rm S}$ and $\GCSG$ to $V_{\rm tg}$ and $V_{\rm TJ}$ (\nameref{sec:methods} and \ref{sec:supp_theory}).

Figure~\ref{Fig4}a plots the results of such calculations in a form of 2D map which shows $R_\mathrm{v}$ dependence on $V_\mathrm{tg}$ and $V_\mathrm{bg}$ (see \nameref{sec:methods} and \ref{sec:supp_theory}). The map captures well all the features of the experiment, in particular, the asymmetric gate voltage dependence of the responsivity and its giant increase when the voltage of top and bottom gates is of the opposite polarity. This is most clearly visible in Fig.~\ref{Fig4}c which compares $R_\mathrm{v}(V_\mathrm {tg})$ dependencies for the cases of zero and finite $V_\mathrm {bg}$. Moreover, our model indicates a broadband character of the tunneling-assisted photoresponse (\ref{Section:Broadband_photoresponse}) as well as provides a remarkable quantitative agreement with experiment provided that $Z_{\rm rad}\approx75~\Omega$, a typical value for the antenna of this type~\cite{BandurinAPL-THz,Bandurin2018}.

The peculiar response of our detector can be understood with the band diagrams shown in Fig.~\ref{Fig4}b. The detector can operate in two regimes: the regime of intraband transport (white, grey, yellow and purple symbols on the map in Fig.~\ref{Fig4}a) and the regime of interband tunneling (green and blue symbols on the map in Fig.~\ref{Fig4}a), depending on the gate voltage configuration. At zero $V_\mathrm {bg}$, BLG is practically gapless, so that the tunnel barrier between the source and the channel is almost absent (white and grey symbols on the map in Fig.~\ref{Fig4}a). In this regime, the device responsivity is controlled by $\RC$ which exhibits a symmetric dependence on $V_\mathrm {tg}$ (cf Fig.~\ref{Fig4}c (black line) and Fig.~2a). On the contrary, when a finite bias is applied to the bottom gate, the tunnel junction is formed as illustrated in Fig.~\ref{Fig4}b (green and blue symbols). Its intrinsic rectifying capability $\RSS$ exceeds that of the  transistor channel $\RC$, as shown in the inset to Fig.~\ref{Fig4}c by several orders of magnitude. This stems from an ultra-strong, exponential sensitivity of the tunnel conductance to the voltage at the junction, as compared to the smooth dependence of $\GCSG$ on $V_\mathrm{tg}$. Moreover, the ac voltage being rectified drops almost completely on a weakly conducting junction but not on the well-conducting channel in the tunneling regime ($\abs{\gamma} \rightarrow 1$). This can be viewed as the 'self-localization' of the ac field in the tunneling rectifier, which contributes to the responsivity enhancement.

Our theory, which successfully describes the response of the BLG device, can also serve to demonstrate the prospects and fundamental limits of TFET-based THz detectors. In the current device, the tunneling is assisted by fluctuations of in-plane electric field induced by local groups of charged impurities~\cite{raikh1987transparency}. In ideal devices, the responsivity would exceed 100 kV/W, according to the model calculations (\ref{Limits}). It is also remarkable that the expected high transconductance of TFET concedes to even higher nonlinearity of the tunnel junction, thus $R_{\rm TJ,i} \gg R_{\rm TG,i}$ in the present and ideal devices. $R_{\rm TG,i}$ can dominate in situations where the source and channel simultaneously possess large gap and remain undoped; thence electron tunneling occurs from a filled valence band of the source to an empty conduction band of the channel in the vicinity of band edges. Large density of states near the bottom of 'Mexican hat'-like spectrum of BLG further increases TFET switching steepness~\cite{our_low-voltage_TFET}. Realization of such band alignment is possible with the application of the drain bias and/or with extra doping gates.

Last, we point to an important advantage of TFET rectifiers with 2D channels, namely, the low internal capacitance of the lateral tunnel junctions. Up to small logarithmic terms, it is given by~\cite{petrosyan1989contact} $C_{\rm TJ} = 2 W \varepsilon \varepsilon_0 \approx 0.4$ fF for the device width of $W = 6.2$ $\mu$m (as in our experiment) and dielectric environment $\varepsilon \approx 4$. The detection cutoff associated with the capacitive shunting of the tunnel junction is therefore expected at $f \sim 1/(2\pi C_{\rm TJ} Z_{\rm rad})\sim5$ THz for $Z_{\rm rad}\approx75~\Omega$ .

%% file: discussion.tex
\section{Discussion}
Despite the fact that our model successfully describes all the features of the observed photoresponse, it does not account for a possible thermoelectric contribution to the responsivity of TFET detectors~\cite{BandurinAPL-THz,viti2020thermoelectric,Vicarelli2012}. Assuming that the Seebeck coefficient varies between $S_{\rm cont}$ in the single-gated region near the source contact and $S_{\rm ch}$ in the double-gated channel, one can estimate the thermoelectric contribution as~\cite{BandurinAPL-THz} $R_{\rm TE} \approx (3 e/2\pi^2 k_B)(S_{\rm cont} - S_{\rm ch})(e|Z_a|/k_B T)(\delta L/L)$, where $\delta L$ is the length of single-gated region and $L$ is the full channel length. Together with Mott's formula for Seebeck coefficient $S = (\pi^2 k_B^2 T/3e) d \ln \sigma/dE_F \sim (\pi^2 k_B^2 T/3eE_F)$ this yields $R_{\rm TE}$ of the order of $5...20$ V/W for Fermi energies in the range $50...200$ meV which is more than two orders of magnitude smaller than the measured responsivity of our TFET detector yet close to that of conventional FET detectors based on gapless monolayer graphene~\cite{BandurinAPL-THz}. Thus, the variations of the 'bulk' thermoelectric parameters cannot explain the observed strong and asymmetric photoresponse of our device.

We note, however, that it is rather challenging to test whether the tunnel junction itself acts as a thermoelectric generator or not~\cite{Tunnel_thermoelectric}. The respective thermoelectric coefficient can be estimated as $S_{\rm TJ} \sim (k_B^2 T/e) d\ln G_{\rm TJ}/d V_\mathrm{g}$, and its functional dependence on the gate voltage would be indistinguishable from the junction rectification described by our model. In principle, measurements of the electron temperature can elucidate the presence of such rectification mechanism, which is however beyond the scope of our work. Nevertheless, even if present, such a mechanism is also due to the presence of the tunnel junction that substantiates the use of TFETs for sensitive THz detection.

Last but not least, we note that reaching the ultimately-low noise-equivalent power would require impedance matching between antenna and TFET~\cite{Bauer_GaNFET_matched}. As the noise power density is proportional to $r_{\rm 2pt}$, and the voltage responsivity saturates at large resistances~\cite{MOM_tunnel_diode}, we can expect the reduction of noise level by a factor of $[r_{\rm 2pt}/Z_\mathrm{rad}]^{1/2}$ in optimized devices. Taking $Z_\mathrm{rad} = 75$ $\Omega$ and $r_{\rm 2pt} = 1...5$ $\mathrm{k\Omega}$ at the points of maximum $R_\mathrm{v}$, we anticipate the ultimate NEP to be $3...8$ times smaller than that reported in Fig.~\ref{Fig3}c and Supplementary Figure~\ref{FigS-comparison}. The simplest way of such matching lies in increasing the device width $W$.

In conclusion, we have shown an opportunity to use TFETs as high-responsivity detectors of sub-THz and THz radiation. Constructing a prototypical device from a BLG dual-gated structure and coupling it to a broadband antenna allowed us to demonstrate the drastic difference between a conventional FET-based approach and TFET-enabled rectification. Furthermore, we have developed a full model enabling one to predict the performance of TFET detectors based on the details of their band structure. This model was applied to the case of BLG TFET detector and successfully captured all the experimentally observed features. As an outlook, we note that BLG is just a convenient platform to demonstrate the performance of TFET-based THz detectors. This approach can be extended to larger-gap materials~\cite{BP-TFET} enabling room-temperature operation, as well as to CMOS-compatible structures~\cite{Si_TFET}. Furthermore, we envision that alternative transistor technologies enabling transconductance beyond Boltzmann limit (phase-change FETs~\cite{Phase_change_FET}, negative capacitance FETs~\cite{Negative_capacitance}) would also demonstrate ultra-sensitive THz detection.

%% file: methods.tex
\section{Methods}
\label{sec:methods}
\subsection{Device fabrication}

To fabricate tunneling-enabled BLG photodetector we first encapsulated BLG between relatively thick hBN crystals using the standard dry-peel technique \cite{Kretinin}. The thickness of hBN crystals was measured by atomic force microscopy. The stack was then transferred on top of a predefined back gate electrode made of graphite deposited onto a low-conductivity THz-transparent silicon wafer capped with a thin oxide layer ($500$ nm). The resulting van der Waals heterostructure was patterned using electron beam lithography to define contact regions. Reactive ion etching was then used to selectively remove the areas unprotected by a lithographic mask, resulting in trenches for depositing electrical leads. Metal contacts to BLG were made by evaporating 3 nm of chromium and 60 nm of gold. Afterwards, a second round of e-beam lithography was used to design the top gate. The graphene channel was finally defined by a third round of e-beam lithography, followed by reactive ion etching using Poly(methyl methacrylate) and gold top gate as the etching mask. Finally, a fourth round of e-beam lithography was used to pattern large bow-tie antenna connected to the source and the top-gate terminals, followed by evaporation of 3 nm of Cr and 400 nm of Au. Antennas were designed to operate at an experimentally relevant frequency range.
 
\subsection{Responsivity measurements}
To perform the photoresponse measurements we used variable temperature optical cryostat equipped with a polyethylene window that allowed us to couple the photodetector to incident THz radiation. A Zytex-106 infrared filter was mounted in the radiation shield of the cryostat to block the 300 K background radiation. The radiation was focused to the bow-tie antenna by a silicon hemispherical lens attached to the silicon side of the chip. The transparency of the silicon wafers to the incident THz radiation over the entire $T-$ range was verified by transmission measurements using a THz spectrometer. Photovoltage was recorded using a home-made data acquisition system based on the PXI-e 6363 DAQ board.

The responsivity of our tunneling-enabled detector was calculated assuming that the full power delivered to the device antenna funnelled into the FET channel. The $R_\mathrm{v}$ determined by this way provides a lower bound for the detectors' responsivity and is usually referred to as extrinsic. The calculation procedure comprised several steps. First the drain-to-source voltage was recorded as a function of the top gate voltage in the dark ($V_\mathrm{dark}$). Then, the dependence of the the drain-to-source voltage, $V_\mathrm{DS}$, on $V_\mathrm{tg}$ was obtained under the illumination with THz radiation. The latter was provided by a calibrated backward wave oscillator generating $=0.13$~THz radiation with the output power $P_\mathrm{out}\approx1~$mW accurately measured using Golay cell. To ensure the characterization of the detector in the linear-response regime, $P_\mathrm{out}$ was further attenuated down to $P_\mathrm{full}\approx2~\mu$W, being the full power delivered to the cryostat window. The difference $V_\mathrm{ph}= V_\mathrm{DS}-V_\mathrm{dark}$ formed the photovoltage.  The responsivity was then calculated as $R_\mathrm{v}=V_\mathrm{ph}/P_{\rm in}$, where $P_{\rm in}\approx P_\mathrm{full}/3.5$ is the power delivered to the antenna after taking into account the losses in the silicon lens and the cryostat optical window ($\approx5.5$ dB). 

In order to study the photoresponse of our detectors at higher $f$, we used a quantum cascade continuous wave laser based on a GaAs/Al$_{0.1}$Ga$_{0.9}$As heterostructure emitting $f=2.026$ THz radiation. Due to the low power of the QCL and non-optimized antenna design at this $f$, the calibration of the delivered to the device antenna power was rather challenging and therefore we only report tunneling-enabled operation of our detector in relative units.    

\begin{figure}[ht!]
	\centering\includegraphics[width=0.9\linewidth]{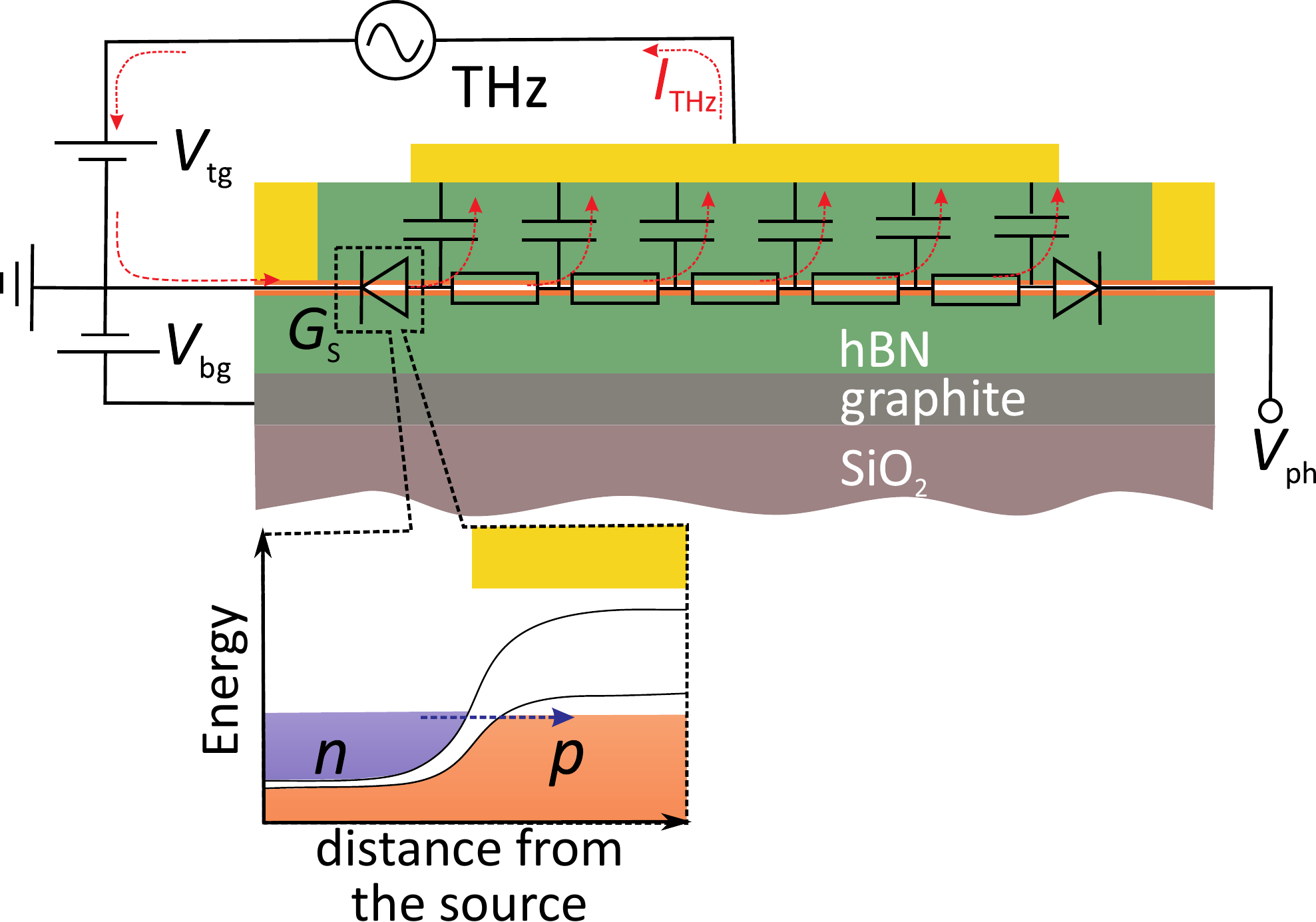}
	\caption{\textbf{Equivalent circuit of the BLG TFET detector}. Antenna is modelled as an equivalent voltage source $V_{\rm ant}$ that generates ac current $I_{\rm THz}$ (red arrows) flowing into the source and escaping the FET channel through the gate capacitance. Rectification occurs mainly at the tunnel barrier between source and channel (see band alignment profile in the inset) with voltage-dependent conductance $G_S$. The doping level and the band gap size is controlled via a simultaneous action of the top and bottom gate voltages, $V_\mathrm{tg}$ and $V_\mathrm{tg}$ respectively. The photovoltage, $V_\mathrm{ph}$, is read out between the source and drain terminals. 
	}
	\label{FigECircuit}
\end{figure}

\subsection{Rectification modelling}
Our detector can be modelled by an equivalent circuit (shown in Fig.~\ref{FigECircuit}) comprising an effective voltage source $V_{\rm ant}$ mimicking an antenna and two nonlinear junctions connected in series with transistor channel. The detector asymmetry, required to obtain a finite photovoltage at zero bias, is provided by the asymmetric connection of antenna between source and gate, and by zero-current condition at the drain. Calculation of detector voltage responsivity $R_{\rm v} = V_{\rm ph}/P_{\rm in}$ includes three distinct steps:
\begin{enumerate}
\item Relating the non-linear $I(V)$ characteristics of circuit elements to the rectified voltage $V_{\rm ph}$.
\item Relating the power incident on antenna with its open-circuit voltage $V_{\rm ant}$.
\item Microscopic calculation of $I(V)$ characteristics for BLG channel and its tunnel contacts.
\end{enumerate}
First, it is convenient to introduce ``voltage-voltage'' responsivity of the TFET, $\RVV = V_{\rm out}/V_{\rm ant}^2$. The responsivity of bare transistor channel coupled to antenna between source and drain is the log-derivative of the dc channel conductance $G_\mathrm{ch}$ with respect to the top gate voltage $V_\mathrm{tg}$~\cite{Sakowicz}, up to a geometrical factor:
\begin{equation}
\label{Resp_channel}
\RC = - \frac{1}{2} \frac{d_b}{d_t + d_b} \frac{\partial \ln G_{C, {\rm dc}}}{\partial V_\mathrm{tg}}.  
\end{equation}
The presence of a tunnel junction with conductance $G_S$ (assumed frequency-independent) depending on the voltage at the junction $V_{\rm TJ}$ and the top gate voltage $V_{\rm tg}$ results in two extra contributions to $\RVV$, which also take the form of log-derivatives: 
\begin{equation}
\label{Resp_tunnel}
\RSS = -\frac{1}{2} \frac{\partial \ln G_S}{\partial V_{\rm TJ}},\qquad \RSG = - \frac{\partial \ln G_S}{\partial V_\mathrm{tg}}.   
\end{equation}

Summation law (\ref{Responsivity_main}) for individual responsivities (\ref{Resp_channel}) and (\ref{Resp_tunnel}) follows directly from Kirchhoff's circuit rules.

At the second stage, the experimentally measured ``voltage-power'' responsivity of the photodetector $\RVP$ is related to the ``voltage-voltage'' responsivity of the TFET as
\begin{equation}
\RVP =  4 Z_{\rm rad} \abs{\frac{Z_{\rm GS}}{Z_{\rm GS} + Z_{\rm rad}}}^2  \RVV,
\end{equation}
assuming the incident radiation is focused within the antenna's effective aperture. The prefactor describes voltage division between the TFET impedance $Z_{\rm GS}$ between gate and source and antenna radiation resistance $Z_{\rm rad}$.

Finally, we calculate the $I(V_{\rm d},V_{\rm tg})$-characteristics of circuit elements microscopically. The FET channel is described within drift-diffusive model with constant mobility $\mu_{\rm BLG} = 10^5$~cm$^2$/(V$\cdot$s), a value close to that found in the experiment. Short junctions are described within quantum ballistic model~\cite{our_low-voltage_TFET}. Both the flux of carriers incident on tunnel barrier and its transparency depend on BLG band structure. This model results in an approximate relation for source junction conductance $G_S \approx \frac{2 e}{\pi^{3/2}\hbar}\Dtun\ktun W$, where $\Dtun$ is the barrier transparency for normal incidence, and $\ktun$ is the characteristic transverse momentum of electrons participating in tunneling. To obtain vanishing junction resistance in the absence of tunnel barrier, $\Dtun$ was replaced by $\Dtun/(1- \Dtun)$~\cite{Landauer_formula}. The appearance of high-transparency regions across the tunnel barrier due to local electric potential fluctuations was modelled as an increase of the average field inside the tunnel barrier $\Ftun$ by a constant value $\Ffluct$. A value of $\Ffluct \approx 8$ kV/cm was extracted from the experimental resistance $R_\mathrm{2pt}$ in the tunneling regime of detector operation.

The calculation of TFET band structure in the double-gated and single-gated regions is based on a parallel-plate capacitor model supplemented with relations between charge densities on graphene layers, their electric potentials, and BLG bandstructure~\cite{BLG_Hamiltonian}. The transient region with tunnel junction was modelled using an original approach, where screening by the charges in BLG was treated approximately by placing a fictitious conducting plane under BLG. The position and potential of this plane are chosen to yield the correct electric potential deep inside the source and channel regions of the BLG. This reduces our electrostatic problem to finding the fringing field of a capacitor, solved analytically by Maxwell~\cite{Maxwell}.

%% file: trivia.tex
\section{Acknowledgements}
This work was supported by the Russian Foundation for Basic Research within Grants No. 18-37-20058 and No. 18-29-20116. Experimental work of IG (photoresponse measurements) was supported by the Russian Foundation for Basic Research (grant 19-32-80028). We acknowledge support of the Russian Science Foundation grant No. 19-72-10156 (NEP analyses) and grant No. 17-72-30036 (transport measurements). The work of GA and  DS (theory of THz detection) was supported by grant \#  16-19-10557 of the Russian Scientific Foundation. K.W. and T.T. acknowledge support from the Elemental Strategy Initiative conducted by the MEXT, Japan, Grant Number JPMXP0112101001, JSPS KAKENHI Grant Number JP20H00354 and the CREST(JPMJCR15F3), JST. D.A.B. acknowledges financial support from Leverhulme Trust. The authors thank  A. Lisauskas, W. Knap, A. I. Berdyugin, Q. Ma and M.S. Shur for helpful discussions.

\section{Data availability}
All data supporting this study and its findings are available within the article and its Supplementary Information or from the corresponding authors upon reasonable request.

\section{Author contributions}
D.S. and D.A.B. conceived the experiment. S.G.X. fabricated devices designed by D.A.B. Photoresponse measurements were carried out by I.G., M.M., and D.A.B. Data analysis was performed by I.G., D.A.B. and D.S. Theory analysis was done by G.A. and D.S. The manuscript was written by D.A.B., G.A. and D.S. with input from I.G. and G.F. Experimental support was provided by I.T., M.M., G.G. and A.K.G.; T.T. and K.W. grew the hBN crystals. D.S., G.F. and D.A.B supervised the project. All authors contributed to discussions.

\section{Competing interests}
The authors declare no competing interests.

%% file: supplementary.tex
\supplementary

\section{Further examples of tunnel-enabled photoresponse} 
\label{Section:Further_examples}
\begin{figure*}[ht!]
\centering\includegraphics[width=0.5\linewidth]{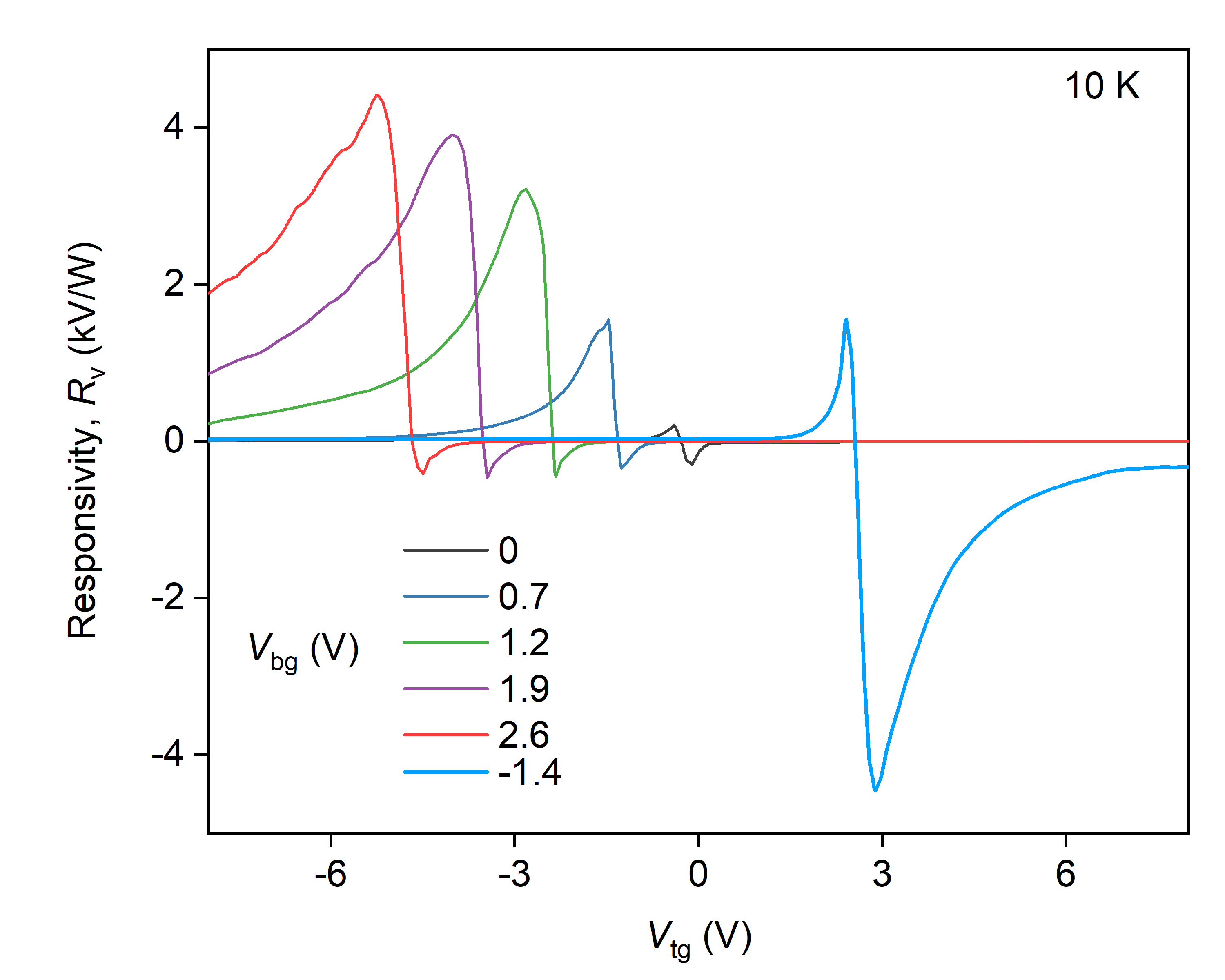}
\caption{\textbf{Tunneling-enabled THz detection.} $R_\mathrm{v}$ as a function of $V_\mathrm{tg}$ for given $V_\mathrm{bg}$ recorded in response to $0.13$~THz radiation. $T=10~$K.}
\label{FigS-Examples}
\end{figure*}

Supplementary Figure~\ref{FigS-Examples} shows further examples of our tunnel detector responsivity $R_\mathrm{v}$ as a function of $V_\mathrm{tg}$ recorded in response to $f=0.13$~THz radiation for varying $V_\mathrm{bg}$. For all $V_\mathrm{bg} \neq 0$, $R_\mathrm{v} (V_\mathrm{tg})$ dependencies are highly asymmetric. With increasing $V_\mathrm{bg}$,  $R_\mathrm{v}$ is increasing and for $V_\mathrm{bg}=2.6$~V reaches 4.5~kV/W overcoming zero $V_\mathrm{bg}$ value by more than an order of magnitude. A similar behaviour was observed if the polarity of $V_\mathrm{bg}$ is reversed (blue curve in Supplementary Figure~\ref{FigS-Examples}). These observations highlights a drastic difference between the field-effect-enabled intraband (black curve) rectification and its interband tunneling counterpart (all other curves).

	\section{Frequency dependence of tunnelling-enabled photoresponse} 
	\label{Section:Broadband_photoresponse}	
	We have also studied the response of our detectors at higher frequency and found consistent tunnelling-enabled highly-asymmetric behaviour when the top and bottom gates are biased with opposite polarity. Examples of $R_\mathrm{v}(V_\mathrm{tg})$ are shown in Supplementary Figure~\ref{FigS-fdep}a for two characteristic $f$ from sub-THz and THz domains. Note, due to the limitation of our measurements (see \nameref{sec:methods}) we only present a relative comparison between $R_\mathrm{v}(V_\mathrm{tg})$ recorded at $f=0.13$~THz and $f=2$~THz. However, our modelling which provides remarkable agreement with experiment at $f=0.13$~THz predicts that TFET detectors are expected to perform equivalently well at both sub-THz and THz frequencies as we show in Supplementary Figures~\ref{FigS-fdep}b,c. 
	
	\begin{figure}[ht!]
	\centering\includegraphics[width=0.95\linewidth]{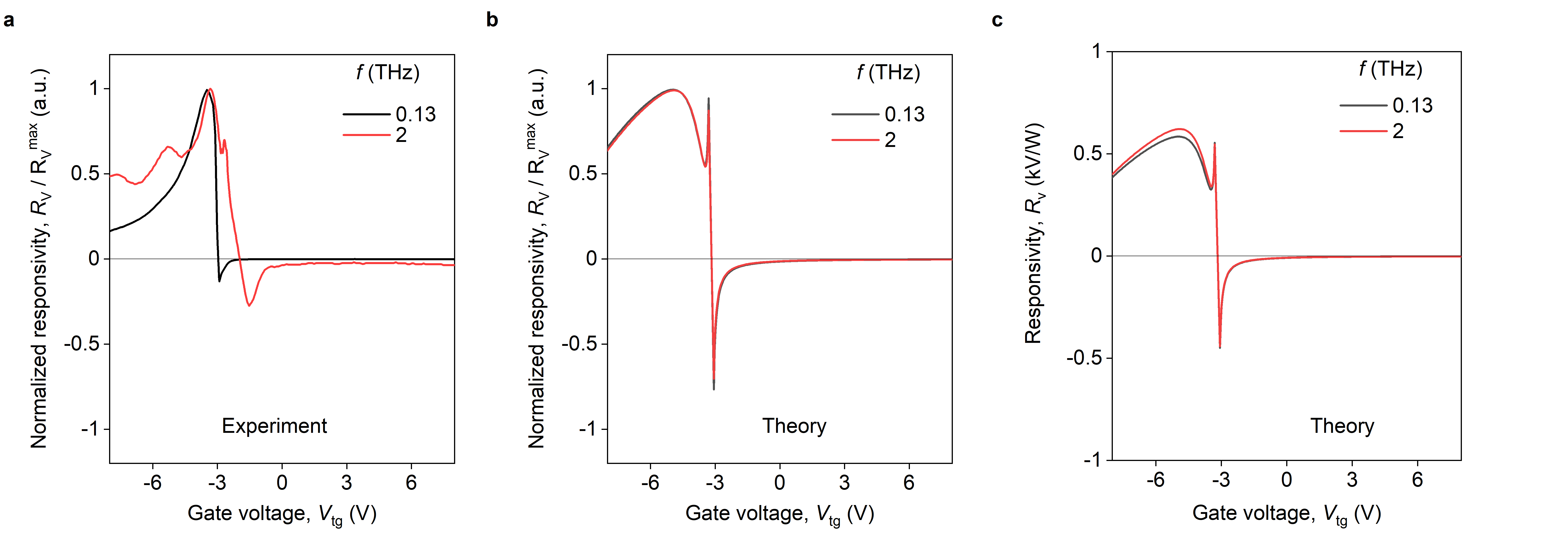}
	\caption{\textbf{Frequency dependence of tunnelling-assisted THz detection.} 
	 \textbf{a} $R_\mathrm{v}$ as a function of $V_\mathrm{tg}$ for $V_\mathrm{bg}=1.2$~V obtained under illumination with THz radiation of given frequency. The data normalized to their maximum value. Peaks in $R_\mathrm{v}$ correspond to the excitation of plasmon-resonances in the detector channel~\cite{supp_our_natcomm}.  \textbf{b, c} Theoretical $R_\mathrm{v}(V_\mathrm{tg})$ dependencies for given $f$: as-calculated (\textbf{c}) and normalized to their maximum value (\textbf{b}).}
	\label{FigS-fdep}
    \end{figure}

\section{Comparison with existing technology} 
\label{Section:Comparison}	
In Supplementary Figure~\ref{FigS-comparison} we compare the performance of our tunnel device with other THz detectors and rectifiers; some of them are available on the market (underlined labels). To this end, we plot their noise equivalent power (NEP) versus temperature, $T$,  at which they operate. The comparison is made for the frequency range $0.1-2$~THz and for the NEP calculated via extrinsic responsivity, i.e. which takes into account the full power delivered to the device. The devices of different types are compared: cooled superconducting bolometers~\cite{supp_scontel,supp_QMC}, cooled semiconducting bolometers~\cite{supp_IRlabs,supp_300mK_bolometers,supp_100mK_bolometer}, kinetic inductance sensors~\cite{supp_TiN_kinetic_inductance,supp_kinetic_inductance_camera}, cooled quantum dot devices~\cite{supp_QD,supp_quantum_rings}, as well as transistor-based detectors~\cite{supp_terasense,supp_graphene_FET_room_T,supp_graphene_FET,supp_our_APL,supp_our_natcomm,supp_AlGaN_HEMT,supp_InGaAs_HEMT,supp_Si-CMOS,supp_cooled_AlGaN_HEMT}, Schottky diodes~\cite{supp_Virginia_Diodes,supp_ACST,supp_cooled_Schottky}, and heterostructure backward diodes~\cite{supp_backward_diode_Zhang,supp_backward_diode_Rahman}. One of the primary tasks for the next-generation THz technology is to produce low-NEP sensors operating at elevated temperatures as indicated by the yellow shaded area in Supplementary Figure~\ref{FigS-comparison}. However, whereas the cooled devices feature exceptionally low NEP, room-$T$ devices are usually characterized by much higher NEP. Our BLG TFET offers a compromise to this enquiry: it features relatively low NEP and operates above liquid helium $T$. Furthermore, our model suggests that TFETs with optimized parameters can feature even lower NEP at room temperature (magenta star in Supplementary Figure~\ref{FigS-comparison}) and thus offer a route to the next-generation THz technology. The details are given in \ref{Limits}.

\begin{figure}[ht!]
	\centering\includegraphics[width=0.5\linewidth]{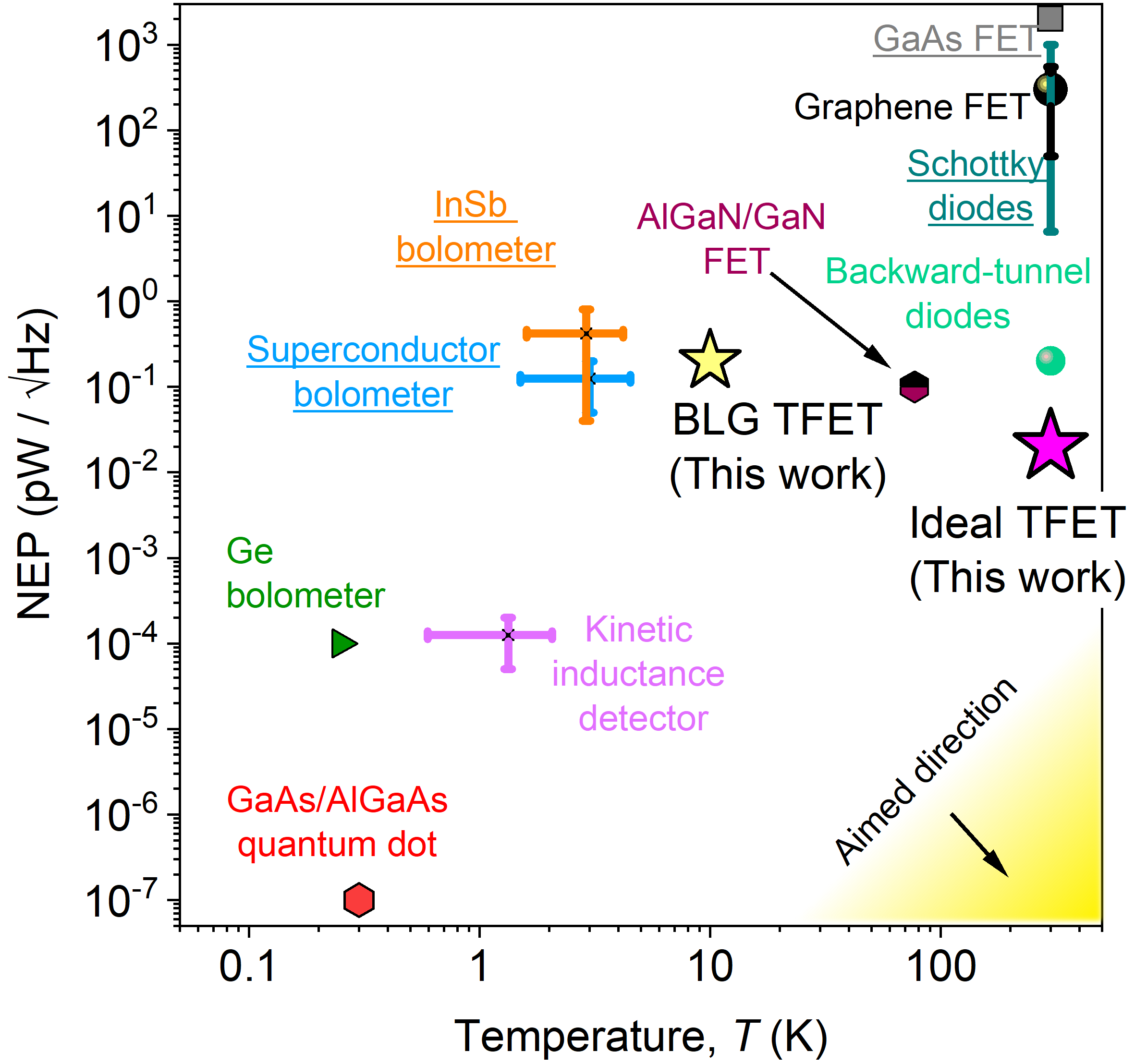}
	\caption{\textbf{Overview of THz detectors.}  NEP for THz detectors of various types plotted against the temperature at which they operate. Vertical error bars represent the spread of the detectors' performance over the frequency range $0.1-2~$THz. Horizontal error bars show the temperature range at which the detectors operate. Underlined labels denote  commercial technology. }
	\label{FigS-comparison}
\end{figure}

\section{Theoretical model of a BLG TFET photodetector}
\label{sec:supp_theory}
\subsection{Modelling of tunneling-assisted THz detection}
\label{sec:responsivity_theory}

	\begin{figure}[ht!]
	\centering\includegraphics[width=0.65\linewidth]{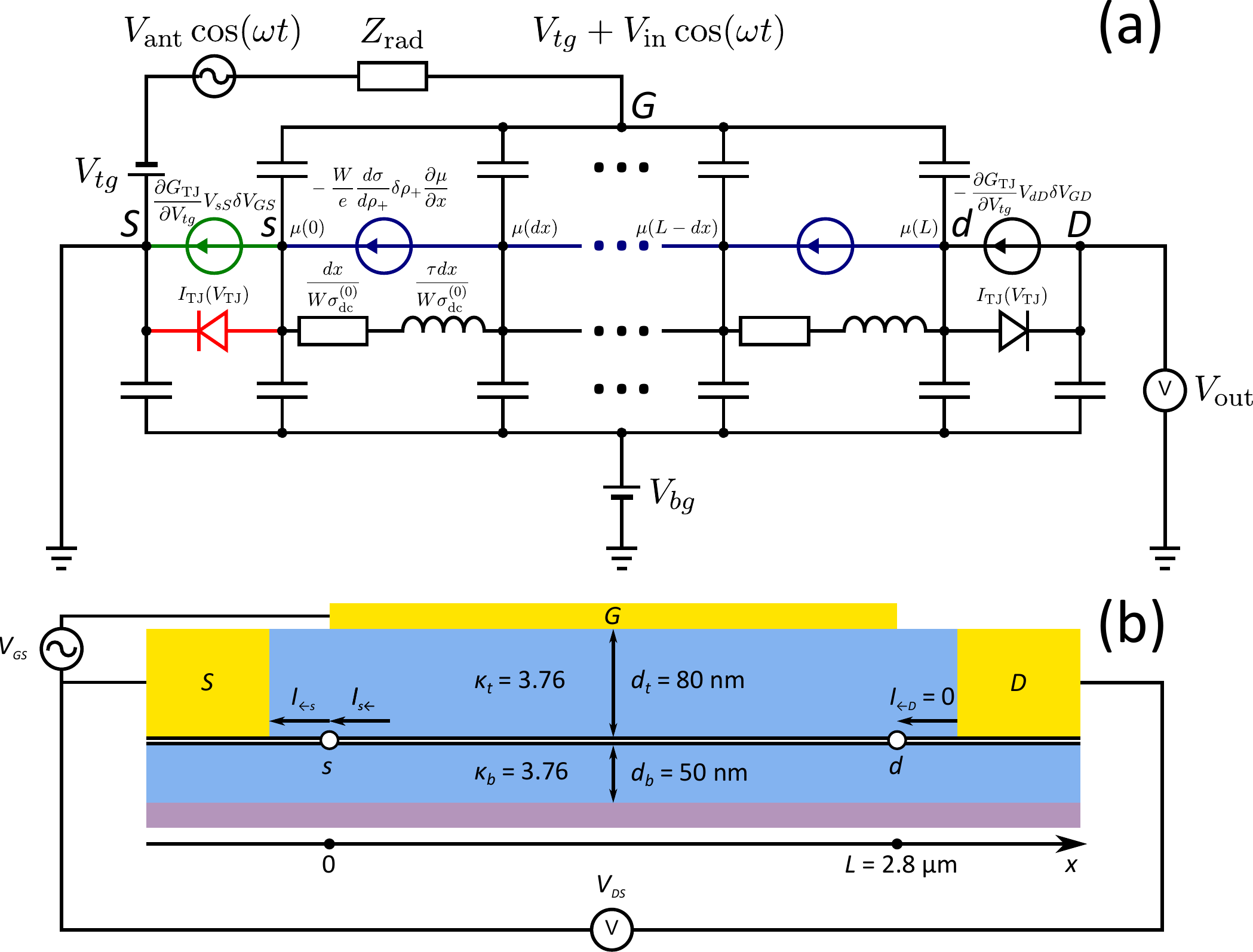}
	\caption{\textbf{Equivalent circuit.} \textbf{a,} Detailed equivalent circuit of a TFET-based detector. Different colors show the origins of different contributions to responsivity: nonlinear current-voltage characteristic of the tunnel junction between source and channel (red), its gate-controlled conductance (represented as an equivalent current source, green), and the gate-controlled channel conductance (represented as a distributed current source, blue). \textbf{b,} Schematic view of our photodetector showing some of the notation used throughout the Supplementary Information.}
	\label{fig:circuit}
    \end{figure}

In this section, we derive a general expression for the responsivity of a TFET. The relevant circuit is shown in Supplementary Figure~\ref{fig:circuit}a. We will treat the TFET as if it was single-gated, since the bottom gate is held at a constant potential and its only function is to open a bandgap in BLG. 

A TFET consists of two rectification units: a tunnel junction between the source and the channel, and the channel itself. (The drain tunnel junction is effectively excluded from the circuit by the zero drain current assumption, at least if the junction is too short to accommodate any spatial inhomogenities of the current.)

When a small ac voltage $V_{{\rm in}}\cos(\omega t)$ is applied between the gate and source, it induces voltages and currents in different parts of the detector, having the general form $(\delta V,\delta I)(t) = \Re \left( (V,I)^{(1)} e^{- i \omega t} \right) + (V,I)^{(2)} + ...\cdot e^{\pm 2i \omega t} + o(V^2_{{\rm in}})$, where we are interested in the first-order and dc second-order components.

We use a non-distributed model for the source junction, meaning current can flow through the junction only in presence of a nonzero voltage drop across the junction and not solely under the action of ac gate voltage. Keeping this is mind, the second-order expansion of its current-voltage characteristic $I_{\leftarrow s}(V_{sS},V_{GS})$ reads

\begin{eq}{source_current}
I^{(1)}_{\leftarrow s} &= G_S(\omega) V^{(1)}_{sS},\\
I^{(2)}_{\leftarrow s} &= G_{S, {\rm dc}} \left[ V^{(2)}_{sS} - \RSS \frac{\abs{ V^{(1)}_{sS} }^2}{2} - \RSG \frac{\Re\left(V^{(1)}_{GS}V^{(1)*}_{sS}\right)}{2} \right],
 \end{eq}
where $G_{S, {\rm dc}}$ and $G_S(\omega)$ are the dc and ac conductance of the junction, $\RSS$ is the intrinsic tunnel junction responsivity, and $\RSG$ is the intrinsic ``tunnel-gate'' responsivity.

When writing similar expressions for the current $I_{s\leftarrow}$ flowing to the source from the channel, we make advantage of the linear dependence between $V^{(1)}_{ds}$ and $V^{(1)}_{Gs}$ arising from zero drain current condition, and use only $V^{(1)}_{Gs}$ and $V^{(2)}_{ds}$ as independent variables (remember that dc gate voltage $V^{(2)}_{Gs}$ does not produce any current by itself):

\begin{eq}{channel_current}
I^{(1)}_{s\leftarrow} &= \GCSG(\omega) V^{(1)}_{Gs},\\
I^{(2)}_{s\leftarrow} &= \GCSDdc \left[ V^{(2)}_{ds} - \RC \frac{\abs{ V^{(1)}_{Gs} }^2}{2} \right],
 \end{eq}
where $\GCSDdc$ is the dc channel conductance, $\GCSG(\omega) \equiv \left(\partial I_{s\leftarrow}(\omega) / \partial V_{Gs}(\omega) \right)|_{I_{\leftarrow d}=0}$ is the ac channel conductance measured between source and gate, and $\RC$ is the intrinsic channel responsivity.

From continuity of current, $I^{(1)}_{\leftarrow s} = I^{(1)}_{s\leftarrow}$ and $I^{(2)}_{\leftarrow s} = I^{(2)}_{s\leftarrow} = I^{(2)}_{\leftarrow D} = 0$, we find that the ac voltage $V^{(1)}_{GS} \equiv V_{{\rm in}}$ applied between the gate and source is divided into voltage $V^{(1)}_{sS}$ at the source tunnel junction and voltage $V^{(1)}_{Gs}$  between the gate and the beginning of the channel:

\begin{eq}{ac_voltages}
V^{(1)}_{sS} &= \frac{\GCSG(\omega)}{G_S(\omega) + \GCSG(\omega)} V_{{\rm in}},\\
V^{(1)}_{Gs} &= \frac{G_S(\omega)}{G_S(\omega) + \GCSG(\omega)} V_{{\rm in}},
 \end{eq}
which are subsequently rectified by the tunnel junction and the channel:

\begin{eq}{dc_voltages}
V^{(2)}_{sS} &= \RSS \frac{\abs{ V^{(1)}_{sS} }^2}{2} + \RSG \frac{\Re\left(V^{(1)}_{GS}V^{(1)*}_{sS}\right)}{2},\\
V^{(2)}_{ds} &= \RC \frac{\abs{ V^{(1)}_{Gs} }^2}{2}.
 \end{eq}

These rectified voltages sum together to yield the output voltage $V^{(2)}_{DS} \equiv V_{{\rm out}}$ of the photodetector (remember that the voltage at the drain junction $V^{(2)}_{Dd} = 0$ because of zero drain current). Total responsivity of the TFET is given by the sum of tunnel junction responsivity, coming from the nonlinear current-voltage characteristic of the source tunnel junction, tunnel-gate responsivity, coming from resistive self-mixing in the gate-controlled source tunnel junction, and channel responsivity, coming from resistive self-mixing in the channel:

\begin{eq}{total_responsivity}
\RVV &\equiv \frac{V_{{\rm out}}}{V^2_{{\rm in}}/2} \equiv \frac{V^{(2)}_{DS}}{\left(V^{(1)}_{GS}\right)^2/2} = \Rdiode + \RTFET + \RFET,\\
\Rdiode &\equiv \abs{ \frac{\GCSG(\omega)}{G_S(\omega) + \GCSG(\omega)} }^2 \RSS,\\
\RTFET &\equiv \Re\frac{\GCSG(\omega)}{G_S(\omega) + \GCSG(\omega)} \RSG,\\
\RFET &\equiv \abs{ \frac{G_S(\omega)}{G_S(\omega) + \GCSG(\omega)} }^2  \RC.
 \end{eq}
 
 We will neglect the frequency dependence of the tunnel junction current-voltage characteristic. With this assumption, intrinsic tunnel junction and tunnel-gate responsivities are given by the logarithmic derivatives of the junction conductance with respect to appropriate voltages:
\begin{eq}{log_derivative_source}
\RSS &= -\frac{1}{2} \left( \frac{\partial \ln G_S}{\partial V_{sS}} \right)_{V_{GS}},\\
\RSG &= - \left(\frac{\partial \ln G_S}{\partial V_{GS}} \right)_{V_{sS}}.
 \end{eq}

Due to the distributed nature of the channel, its current-voltage characteristics are inherently frequency-dependent. Nevertheless, the intrinsic channel responsivity can also be expressed in terms of the logarithmic derivative of its dc conductance, see \ref{sec:channel_responsivity}:
\begin{eq}{log_derivative_channel}
\RC \approx - \frac{1}{2} \frac{d_b}{d_t + d_b} \left(\frac{\partial \ln \GCSDdc}{\partial V_{Gs}}\right)_{V_{ds}=0}.
 \end{eq}
A similar expression was originally derived in \Cite{supp_Sakowicz} for a single-gated FET. The extra prefactor represents the gate voltage division in a double-gated structure with top and bottom gate dielectrics of thicknesses $d_t, d_b$.

TFET responsivity \eqref{total_responsivity} describes its response to the ac voltage at the gate, while the experimentally measured photodetector responsivity $\RVP$ describes response to the power $P_{{\rm in}}$ incident on the antenna. The relation between these responsivities can be obtained by considering the complete circuit of the photodetector, including the antenna radiation resistance $Z_{\rm rad}(\omega)$ (Supplementary Figure~\ref{fig:circuit}a). Assuming the incident radiation is focused within the antenna's effective aperture, the incident power can be converted into the effective voltage $V_{\rm ant} = \sqrt{8 Z_{\rm rad}(\omega) P_{{\rm in}}}$~\cite{supp_recieved_power_definition}, which is divided between $Z_{\rm rad}(\omega)$ and the TFET gate-to-source ac impedance
\begin{eq}{TFET_impedance}
Z_{\rm GS}(\omega) = G^{-1}_S + \GCSG^{-1}(\omega),
\end{eq}
yielding
\begin{eq}{voltage_to_power_responsivity}
\RVP \equiv \frac{V_{{\rm out}}}{P_{{\rm in}}} = 8 Z_{\rm rad}(\omega) \frac{V_{{\rm out}}}{\abs{V_{\rm ant}}^2} =  4 Z_{\rm rad}(\omega) \abs{\frac{Z_{\rm GS}(\omega)}{Z_{\rm GS}(\omega) + Z_{\rm rad}(\omega)}}^2  \RVV.
\end{eq}

\subsection{Bandstructure and charge density in bilayer graphene}
\label{sec:bandstructure}
\begin{fig}[t]{bands}{supp_figs/supp_bands}
\textbf{Bilayer graphene bandstructure.} Bandstructure of biased bilayer graphene described by Hamiltonian \eqref{Hamiltonian} (only the conduction and valence bands are shown). Circular band extrema are highlighted in yellow.
\end{fig}

BLG in external electric field is described by the Hamiltonian~\cite{supp_BLG_Hamiltonian, supp_BLG_handbook}
\begin{eq}{Hamiltonian}
\hat H\left( \vec{k} \right) = \left( {\begin{array}{*{20}c}
   {- e \phit} & {\hbar v_0(\pm k_x - i k_y )} & 0 & 0  \\
   {\hbar v_0(\pm k_x + i k_y )} & {- e \phit} & {\gamma _1 } & 0  \\
  0 &  {\gamma_1 } & {- e \phib} & {\hbar v_0(\pm k_x - i k_y )}  \\
   0 & 0 & {\hbar v_0(\pm k_x + i k_y )} & {- e \phib}  \\
\end{array}} \right),
\end{eq}
in the vicinity of $K, K'$ points of the Brillouin zone, where $\phit$, $\phib$ are the electric potentials at top and bottom graphene layers, $\gamma_1 = 0.38$ eV, $v_0 = 10^6$ m/s, and the signs depend on the valley.

The corresponding conduction and valence band dispersions are
\begin{eq}{dispersion}
E_{c,v}(k) &= - e\phip \pm E(k),\\
E(k) &= \sqrt{ \frac{E_g^2}{4} + \left[ \sqrt{\frac{\gamma_1^2 - E_g^2}{4} + \left( \hbar v_0 k \right)^2} - \frac{\gamma_1^2}{2\sqrt{\gamma_1^2 - E_g^2}} \right]^2 }
\end{eq}
with a bandgap
\begin{eq}{bandgap}
E_g\left(\phim\right) = \frac{\gamma_1}{\sqrt{\gamma_1^2 + e^2 \phim^2}}\abs{e\phim},
\end{eq}
where $\phip \equiv (\phit + \phib)/2$ is the average potential of graphene layers, and $\phim \equiv \phit - \phib$ is the interlayer voltage. The bands have a ``Mexican hat'' shape with circular extrema around the corners of the Brillouin zone (Supplementary Figure~\ref{fig:bands}).

The inverse dispersion relation is 
\begin{eq}{inverse_dispersion}
k_{\pm}(E - e\phip) = \frac{1}{\hbar v_0}\sqrt{E^2 + \frac{e^2 \phim^2}{4} \pm \sqrt{\left( \gamma_1^2 + e^2 \phim^2 \right)\left( E^2 - \frac{E_g^2}{4} \right)}}.
 \end{eq}
It is double-valued within the ``Mexican hat'' region, $E_g/2 < \abs{E} \leq \abs{e \phim}$, while only a single solution $k_+$ remains above the hat, $\abs{E} > \abs{e \phim}$.

Given the dispersion relation \eqref{dispersion}, we can express the charge density $\rho_+ \equiv \rho_t + \rho_b$ in BLG at zero temperature through the chemical potential measured from the midgap, $\mutld \equiv \mu + e\phip$, and vice versa:

\begin{eq}{zero-T_charge_density}
\rhop \left( \mutld \right) &= \begin{cases}
0 & \text{if } \abs{\mutld} < \frac{E_g}{2},\\
-e\frac{k_F^2}{\pi} \sgn\mutld, k_F = k_{+}\left(\mutld\right) & \text{if } \abs{\mutld} \geq \frac{\abs{e\phim}}{2},\\
-e\frac{k_{F+}^2 - k_{F-}^2}{\pi} \sgn\mutld, k_{F\pm} = k_{\pm}\left(\mutld\right) & \text{if } \frac{E_g}{2} < \abs{\mutld} < \frac{\abs{e\phim}}{2},
\end{cases}\\
\mutld \left( \rhop \right) &= \begin{cases}
- E\left(k_F\right) \sgn\rhop, k_F = \sqrt{\frac{\pi \abs{\rhop}}{e}} & \text{if } \abs{\rhop} \geq \rhohat,\\
- \sqrt{\frac{E_g^2}{4} + \frac{1}{4 \left( \gamma_1^2 + e^2 \phim^2 \right)} \left(\hbar v_0 \sqrt{\frac{\pi \abs{\rhop}}{e}} \right)^4} \sgn\rhop & \text{if } 0 < \abs{\rhop} < \rhohat\\
\end{cases}
\end{eq}
where $\rhohat = (e/\pi)(e\phim/\hbar v_0)^2$ is the charge density corresponding to $\mutld = \pm e \phim/2$ (the Fermi level positioned at the tip of the ``Mexican hat''). We have taken into account the double valley degeneracy in BLG.

\subsection{Electrostatics of double-gated bilayer graphene}
\label{sec:electrostatics}
\begin{widefig}{capacitance}{supp_figs/supp_capacitance}
\textbf{Accuracy of the constant interlayer quantum capacitance approximation.} Ratio between the interlayer charge transfer calculated using a constant interlayer quantum capacitance $\Cmq = 3e^2 \gamma_1 / (4 \pi \hbar^2 v^2_0)$ and the exact interlayer charge transfer $\rhom$ calculated from Hamiltonian \eqref{Hamiltonian} as described in~\Cite{supp_BLG_Hamiltonian}. Left panel: zero temperature, right panel: $T = 77$~K.
\end{widefig}

To calculate the band diagram of our TFET, we seek approximate analytical solution of electrostatic equations for double-gated BLG.

Let $\Vtg$, $\Vbg$ be the potentials of the top and bottom gate, $d_t$, $d_b$ the thicknesses of dielectric layers separating BLG from the gates, $\kappa_t$, $\kappa_b$ the dielectric constants of these layers, and $d$ the interlayer distance in BLG. Then, the total charge density $\rho_+ \equiv \rho_t + \rho_b$ and interlayer charge transfer $\rhom \equiv(\rho_t - \rho_b)/2$ are related to the electric potentials $\phit$, $\phib$ of top and bottom graphene layers  by

\begin{eq}{exact_electrostatics}
\rhop &= - \Ct \left( \Vtg - \phit \right) - \Cb \left(\Vbg - \phib \right),\\
\rhom &= -\frac{\Ct \left( \Vtg - \phit \right)}{2} + \frac{\Cb \left(\Vbg - \phib \right)}{2} + \Cmcl\left(\phit - \phib \right),
 \end{eq}
where we introduced capacitances per unit area $\Cmcl \equiv \eo/d$, $\Ct \equiv \kappa_t \eo / d_t$, $\Cb \equiv \kappa_b \eo / d_b$.

The potentials of graphene layers stay close to the Fermi level (compared to the gate voltages), and we can substitute $V_{t/b} - \varphi_{t/b} \rightarrow V_{t/b} + \mu/e$ in the second equation. We cannot do the same in the first equation, otherwise it would not work properly in the undoped case. Instead, in the first equation we approximate $\phit \approx \phib \approx \phip$ to decouple $\rhop$, $\phip$ and $\rhom$, $\phim$:

\begin{eq}{approx_electrostatics}
\rhop &\approx - \Ct \Vtg - \Cb \Vbg + \left( \Ct + \Cb \right)\phip,\\
\rhom &= -\frac{\Ct \left( \Vtg + \mu/e \right)}{2} + \frac{\Cb \left(\Vbg + \mu/e \right)}{2} + \Cmcl\phim.
 \end{eq}
The introduced approximations essentially amount to a minor shift of gate voltages, by the order of magnitude equal to $\phit$, $\phib$.

These equations have to be supplemented with explicit expressions for $\rho_{\pm}(\mutld,\phim)$ in BLG ($\mutld \equiv \mu + e\phip$ is the chemical potential with respect to the midgap). To facilitate analytical treatment, we use zero-temperature expression for the total charge density \eqref{zero-T_charge_density} and a constant quantum capacitance model for the interlayer charge transfer:

\begin{eq}{charge_densities}
   \rhop &= \begin{cases}
      0 & \text{if } \abs{\mutld} \leq \frac{E_g\left(\phim\right)}{2}, \\
      \rhop\left( \mutld \right) & \text{if } \abs{\mutld} > \frac{E_g\left(\phim\right)}{2},
    \end{cases}\\
\rhom&\approx -\Cmq \phim.
 \end{eq}
The constant interlayer quantum capacitance $\Cmq = 3e^2 \gamma_1 / (4 \pi \hbar^2 v^2_0)$ approximates the interlayer charge transfer in BLG over a wide range of bangaps and doping levels within 50\% accuracy (see Supplementary Figure~\ref{fig:capacitance}).

Now, the equation for $\rhom$ becomes trivial to solve, while the equation for $\rhop$ requires some additional simplifications to allow analytical solution. We consider two opposite cases: (1) Fermi level lies within the bandgap, (2) Fermi level lies outside the gap. In the first case, $\rho_+ = 0$ at zero temperature, and $\phip$ is readily obtained from \eqref{approx_electrostatics}. In the second case, we can pick some initial guess for $\phip$, find $\rho_+$ from \eqref{approx_electrostatics}, and find a better approximation for $\phip$ from \eqref{charge_densities}. Since the quantum capacitance $\Cpq \sim \eo/d$ is much larger than $\Ct + \Cb$, the precise value of the initial guess is unimportant, and we initially assume the Fermi level is pinned at the band edge, $\mutld = \pm E_g/2$ (this choice avoids spurious discontinuities in $\phip(\Vtg,\Vbg)$).

The overall procedure is summarized in the following equations:

\begin{eq}{analytical_electrostatics}
\phim &\approx \frac{\Ct \left( \Vtg + \mu/e \right) - \Cb \left( \Vbg + \mu/e \right)}{2 \Cm}, \Cm \equiv \Cmcl + \Cmq,\\
E_g &= E_g\left( \phim \right),\\
\mutldo &= e\frac{\Ct \left( \Vtg + \mu/e \right) + \Cb \left( \Vbg + \mu/e \right)}{\Ct + \Cb},\\
\mutld &\approx \begin{cases}
\mutldo & \text{if } \abs{\mutldo} \leq \frac{E_g}{2}, \\
\mutld\left( \rhop \right), \rhop = - \frac{\Ct + \Cb}{e} \left( \mutldo - \frac{E_g}{2} \sgn \mutldo \right)  & \text{if } \abs{\mutldo} > \frac{E_g}{2},
    \end{cases}
 \end{eq}
where $E_g\left( \phim \right)$, $\mutld\left( \rhop \right)$ are given by \eqref{bandgap} and \eqref{zero-T_charge_density}.

In our calculations, we used $d_t = 80$~nm, $d_b = 50$~nm, $d = 0.335$~nm, and $\kappa_t = \kappa_b = 3.76$ (out-of-plane static dielectric constant of hexagonal boron nitride~\cite{supp_hBN_dielectric_constant}).

We use this parallel-plate capacitor model to find the electric potentials $\phipS$, $\phipC$ and interlayer voltages $\phimS$, $\phimC$ in the source region and in the channel in absence of ac signal, and also to calculate the channel response to an ac signal, see \ref{sec:channel_responsivity}. In the source region, there is only the bottom gate, while the role of a top gate is played by infinity, held at zero potential. This means $\Ct = 0$, and the top gate disappears from the equations.

\subsection{Tunneling field}
\label{sec:field}

	\begin{figure}[t]
	\centering\includegraphics[width=0.35\linewidth]{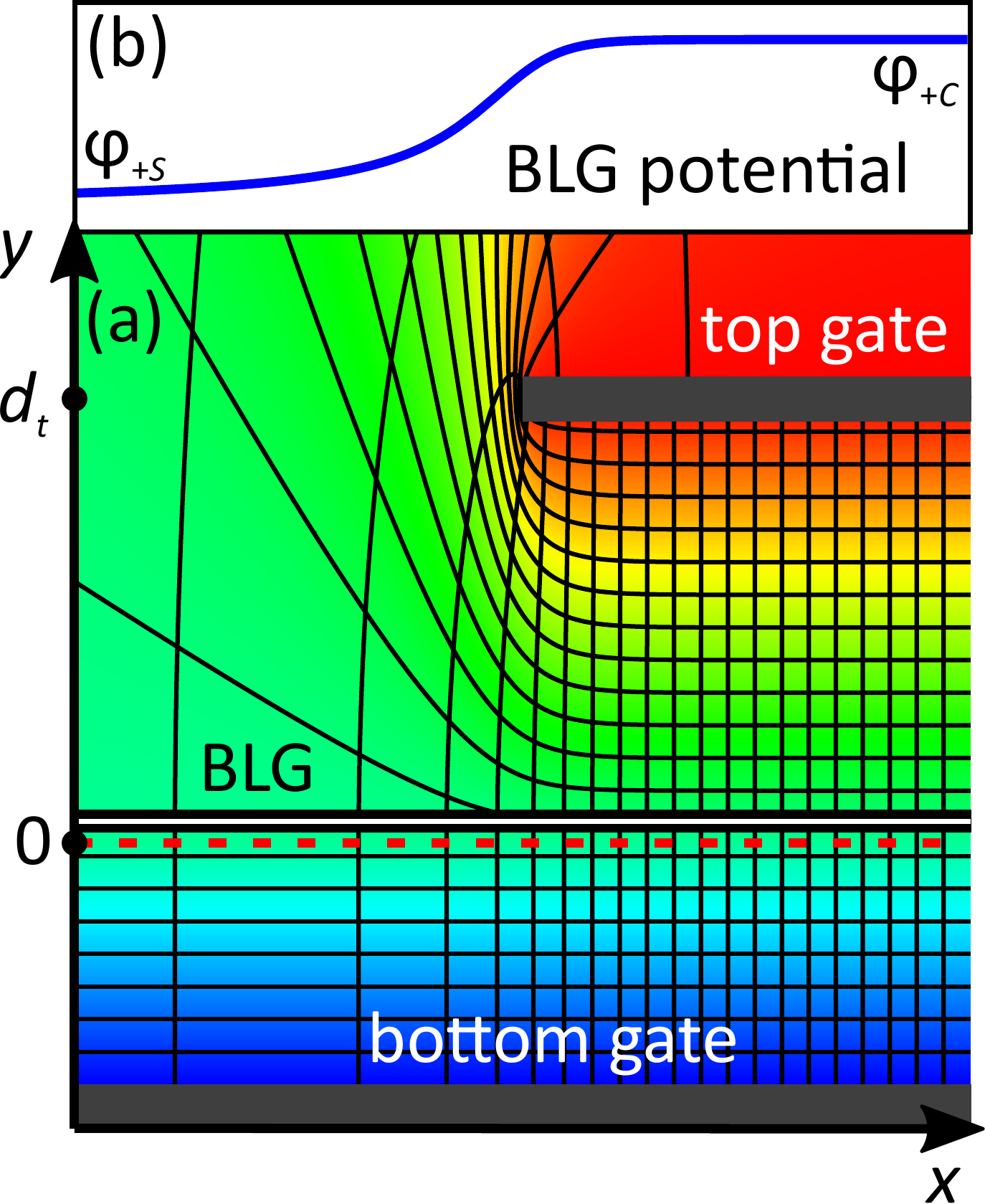}
	\caption{\textbf{Electric field in the tunnel junction.} \textbf{a} Color map showing the distribution of electric potential near the source-channel junction. Black lines: equipotential lines and field lines. Dashed red line: fictitious conductor introduced to obtain the correct potential in the source and the channel without explicitly considering screening by BLG. Potential above the fictitious conductor was calculated as prescribed in \Cite{supp_Maxwell}, potential below the fictitious conductor was calculated in the parallel-plate capacitor model. \textbf{b} Electric potential $\phip(x)$ inside BLG.}
	\label{fig:field}
    \end{figure}
    
In a TFET based on double-gated BLG, a tunnel juntion is formed under the top gate edge, where the parallel-plate capacitor model of \ref{sec:electrostatics} cannot be applied, and an accurate calculation of the tunneling field requires solving a two-dimensional electrostatic problem. This problem can be solved analytically in the absence of BLG~\cite{supp_Maxwell}, and the answer is

\begin{eq}{Maxwell_field}
\Ex = \frac{1 - \phitld}{\left[ 1 + \ytld \left( 1 - \phitld \right) \cot\ytld\phitld \right]^2 + \ytld^2 \left( 1 - \phitld \right)^2},
 \end{eq}
where

\begin{eq}{dimensionless_field_vars}
\Ex \equiv \frac{E_x d_{tb}}{\pi \left[ \varphi\left(x=-\infty\right) - \varphi\left(x=+\infty\right) \right]}, \ytld \equiv \frac{\pi y}{d_{tb}}, \phitld \equiv \frac{\varphi - \varphi\left(x=-\infty\right)}{\varphi\left(x=+\infty\right) - \varphi\left(x=-\infty\right)}
 \end{eq}
are the dimensionless field in the plane of BLG, dimensionless position of BLG with respect to the gates ($\ytld = 0$ at the bottom gate and $\pi$ at the top gate), and dimensionless electric potential at the point where the field is calculated. $d_{tb} = d_t + d_b + d$ is the distance between gates,  $\varphi\left(x=-\infty\right) = \Vbg$ and $\varphi\left(x=+\infty\right) = \Vbg+(y/d_{tb})(\Vtg-\Vbg)$ are the electric potential in the source region and in the channel. The top and bottom dielectrics are assumed to be the same, as in our experiment.

Across a wide range of $\ytld$, $\Ex$ is close to its low-$\ytld$ limit

\begin{eq}{approx_Maxwell_field}
\Ex \approx \phitld^2 \left( 1 - \phitld \right).
 \end{eq}

In the presence of BLG, exact calculation of the tunneling field would require solving the two-dimensional electrostatic problem numerically. To avoid this, we notice that adding BLG into the system reduces $\varphi\left(x=+\infty\right) - \varphi\left(x=-\infty\right)$  from several volts to tens or hundreds of millivolts. This suggests to approximate the screening by BLG via introducing a fictitious perfect conductor placed very close to the BLG. The potential of this conductor and its distance from the BLG are chosen so as to reproduce the correct potentials in the source and channel regions of BLG.

The resulting electric potential distribution in the system is shown in Supplementary Figure~\ref{fig:field}. Introducing the fictitious conductor allows us to keep \Eqref{approx_Maxwell_field} for the electric field in BLG, if $d_{tb}$ is replaced with $d_t$ in the definition of  (\Eqref{dimensionless_field_vars}), and $\varphi\left(x=\pm\infty\right)$ are calculated in the parallel-plate capacitor model described in \ref{sec:electrostatics}.

Knowing the distribution of electric potential in BLG, we can calculate the tunnel current though the source-channel junction. Before we actually do this, we introduce two additional simplifications. First, we neglect field variations inside the barrier and assume tunneling through uniform field. This field is calculated at the point where the tunneling electron crosses the midgap ($E + e\varphi = 0$, where $E$ is the electron energy). Second, instead of using different values of the tunneling field for electrons of different energies, we use a single value calculated for energy $E = (\Etunmin + \Etunmax)/2$. $\Etunmin$ and $\Etunmax$ are the boundaries of the energy region where tunneling is possible. Assuming zero temperature and both quasi-Femi levels $\mu_S$, $\mu_C$ in the source and the channel (near its beginning) lying within the band overlap region, we can write $\Etunmin = \min\left\{ \mu_S, \mu_C \right\}$ and $\Etunmax = \max\left\{ \mu_S, \mu_C \right\}$. (Remember that we are interested in the small-signal case, when the quasi-Fermi levels are close to each other and either lie both inside the band overlap region, or both outside. In the latter case, tunneling is impossible.)

To summarize, we use the following expression for the tunneling field:
\begin{eq}{tunneling_field}
\Ftun \approx \frac{\pi \abs{\phipC - \phiptun}}{d_t} \left(\frac{\phiptun - \phipS}{\phipC - \phipS}\right)^2,
 \end{eq}
where $\phipS$, $\phipC$ are calculated as described in \ref{sec:electrostatics}, and $-e\phiptun = \left(\mu_S + \mu_C\right)/2$.

\subsection{Responsivity of the source-channel junction}
\label{sec:tunnel_responsivity}

A zero-temperature ballistic expression for the tunnel current through the source-channel junction is
\begin{eq}{exact_tunnel_current}
\Jtun = 8 e W \int_{\mu_C}^{\mu_S} \frac{dE}{2\pi\hbar} \int_{-\kmax(E)}^{\kmax(E)} \frac{d\kperp}{2\pi} \calD\left(E, \kperp\right)
 \end{eq}
 if $\mu_S > \mu_C$ (the opposite case is treated similarly). Here, $W = 6.2$~$\mu$m is the channel width, $\calD\left(E, \kperp\right)$ is the barrier transparency, the wavevector integral is taken up to the maximum possible transverse wavevector $\kmax(E)$ that an electron with energy $E$ can have both in the source and in the channel, and the factor of 8 results from two spin projections, two valleys, and two tunneling paths in the imaginary $k$-space (interference between them~\cite{supp_tunneling_paths_interference} is neglected).
 
An analytical approximation can be derived by expanding the WKB barrier transparency in powers of $\kperp$ up to second order and extending the wavevector integration up to infinity~\cite{supp_our_low-voltage_TFET}:
\begin{eq}{approx_tunnel_current}
\Jtun &\approx \frac{2 e}{\pi^{3/2}\hbar}\Dtun\ktun W \left( \mu_S - \mu_C \right),\\
\Dtun &\approx \exp\left( -\frac{\pi\sqrt{\gamma_1 \Egtun^3}}{4\hbar v_0 e \Ftun} \right),\\
\ktun &\approx \sqrt{\frac{4}{\pi} \sqrt{\frac{\gamma_1}{\Egtun}} \frac{e \Ftun}{\hbar v_0}}.
 \end{eq}
We assume that the transition between the source and the channel has the same shape for both the interlayer voltage and the electric potential and, therefore, the tunnel current flows through the bandgap $\Egtun \approx \abs{\phimtun}$, where
\begin{eq}{tunnel_gap}
\frac{\phimtun - \phimS}{\phimC - \phimS} = \frac{\phiptun - \phipS}{\phipC - \phipS}.
 \end{eq}
At experimental conditions, the bandgap does not exceed 60 meV, so we use $E_g \approx \abs{\phim}$ instead of a more accurate expression \eqref{bandgap}.

Expressions \eqref{exact_tunnel_current}, \eqref{approx_tunnel_current} require that the chemical potentials $\mu_S$, $\mu_C$ are taken at the points where the deviations of the carrier distributions from the Fermi-Dirac form are negligible. Since we consider the tunnel junction connected in series with the channel, we need an expression for the tunnel current in terms of the voltage \emph{directly at the junction}, otherwise a certain part of the channel would be counted twice. This can be achieved by introducing a $1 - \Dtun$ correction in the denominator:
\begin{eq}{tunnel_current_Landauer_correction}
\Jtun &\approx \frac{2 e}{\pi^{3/2}\hbar}\frac{\Dtun}{1 - \Dtun}\ktun W \left( \mu_S - \mu_C \right),\\
 \end{eq}
similarly to the one-dimensional Landauer formula containing $\calD/(1 - \calD)$~\cite{supp_Landauer_formula,Note1}.

The idealistic model that led to \Eqref{tunnel_current_Landauer_correction} gives very small barrier transparency and huge tunnel resistance, orders of magnitude larger than in our experiment. This suggests there is some mechanism affecting the junction resistance, most likely electron-hole puddles, that create field fluctuations and may increase the average tunneling field. We take this effect into account phenomenologically, introducing a single fitting parameter $\Ffluct$, which represents the average fluctuating field and is added to the tunneling field \eqref{tunneling_field} calculated without disorder:
\begin{eq}{tunnel_current_with_puddles}
\Jtun &\approx \frac{2 e}{\pi^{3/2}\hbar}\frac{\Dtun}{1 - \Dtun}\ktun W \left( \mu_S - \mu_C \right),\\
\Dtun &\approx \exp\left( -\frac{\pi\sqrt{\gamma_1 \Egtun^3}}{4\hbar v_0 e \left( \Ftun + \Ffluct \right)} \right),\\
\ktun &\approx \sqrt{\frac{4}{\pi} \sqrt{\frac{\gamma_1}{\Egtun}} \frac{e \left( \Ftun + \Ffluct \right)}{\hbar v_0}}.
 \end{eq}
This is the final expression for the tunnel current that we used in our calculations. The value $\Ffluct = 8$~kV/cm was found by fitting the experimental resistance in the tunnel regime and simultaneously gave responsivity in reasonable agreement with the experiment.

Assuming grounded source, $\mu_S = 0$, we identify $\mu_C$ with $-e V_{sS}$ and $\Vtg$ with $V_{GS}$ from \ref{sec:responsivity_theory}. Now, we can calculate the junction conductance as $G_S = -e \partial \Jtun / \partial \mu_C$ and the intrinsic tunnel junction and tunnel-gate responsivities through \eqref{log_derivative_source}. When the doping types of source and channel are the same, or channel is undoped, there is no \emph{tunnel} junction. In this case, we set the junction conductance to infinity and tunnel junction and tunnel-gate responsivities to zero.

\subsection{Responsivity of a long double-gated channel}
\label{sec:channel_responsivity}

In this section, we consider resistive self-mixing in a long~\cite{Note2} \emph{double-gated} channel and find its responsivity. Our derivation closely follows that of \Cite{supp_Sakowicz}, but extends it by (1) allowing the carrier density to depend separately on the top gate voltage and the Fermi level (because $\rhop = \rhop\left(\mu + e\Vtg\right)$ is no longer true in the presence of a bottom gate), (2) using frequency-dependent channel conductivity.

The basic assumptions of our model are that the dc channel conductivity $\sigmadc(x,t)$ is instantaneously related to the local charge density $\rhop(x,t)$, which, in turn, is related (also locally and instantaneously) to the top gate voltage $\Vtg(t)$ and the Fermi level $\mu(x,t)$. Response to ac perturbations is described within the Drude model. Together with the charge conservation, we get a system of four equations:

\begin{eq}{channel_equations}
\rhop(x,t) &= \rhop\left(\Vtg (t), \mu(x,t)\right),\\
\frac{\partial \Jloc(x,t)}{\partial t} &= -\frac{1}{e} \frac{\sigmadc(x,t)}{\tau} \frac{\partial \mu(x,t)}{\partial x} - \frac{\Jloc(x,t)}{\tau},\\
\sigmadc(x,t) &= \sigmadc\left[\rhop(x,t)\right],\\
\frac{\partial \rhop(x,t)}{\partial t} &= \frac{\partial \Jloc(x,t)}{\partial x},
 \end{eq}
with the boundary conditions of grounded source and zero drain current:

\begin{eq}{channel_boundary_conditions}
\mu(0,t) &= 0, \Jloc(+\infty,t) = 0.
 \end{eq}
The top gate voltage consists of a constant bias and an ac signal, $\Vtg(t) = \Vtg^{(0)} + \Vin \cos\left( \omega t \right)$. (Hereafter, quantities in the absence of the ac signal will be denoted by the $^{(0)}$ superscript, while next orders in $\Vin$ will be denoted by $^{(1)}$ and $^{(2)}$, as in \ref{sec:responsivity_theory}.)

To the first order in $\Vin$, we obtain

\begin{eq}{first-order_channel_equations}
\rhop^{(1)} &= \frac{\partial \rhop}{\partial \Vtg} \Vin + \frac{\partial \rhop}{\partial \mu} \mu^{(1)},\\
\Jloc^{(1)} &= -\frac{1}{e} \frac{\sigmadc^{(0)}}{1 - i\omega\tau} \frac{\partial \mu^{(1)}}{\partial x},\\
\sigmadc^{(1)} &= \frac{d \sigmadc}{d \rhop} \rhop^{(1)},\\
-i \omega \rhop^{(1)} &= \frac{\partial \Jloc^{(1)}}{\partial x}.\\
 \end{eq}
Using the boundary conditions \eqref{channel_boundary_conditions}, we get the following solution

\begin{eq}{first-order_channel_solutions}
\mu^{(1)}(x) &= -\left( \frac{\partial \rhop}{\partial \mu} \right)^{-1}\frac{\partial \rhop}{\partial \Vtg} \Vin \left[ 1 - e^{i \qpl x} \right],\\
\rhop^{(1)}(x) &= \frac{\partial \rhop}{\partial \Vtg} \Vin  e^{i \qpl x},\\
\Jloc^{(1)}(x) &= -\frac{1}{e} \frac{\sigmadc^{(0)}}{1 - i\omega\tau} i \qpl \left( \frac{\partial \rhop}{\partial \mu} \right)^{-1}\frac{\partial \rhop}{\partial \Vtg} \Vin e^{i \qpl x},\\
\qpl &\equiv \sqrt{\frac{i \omega (1 - i\omega\tau)}{\sigmadc^{(0)}} \frac{\partial \rhop}{\partial \left( -\mu/e \right)} }.
 \end{eq}

Having found the first-order current,  we can write the channel ``source-gate'' conductance ($W$ is the channel width):

\begin{eq}{open-drain_transconductance}
\GCSG(\omega) \equiv \frac{\Jloc^{(1)}(0)W}{\Vin} = - \frac{\sigmadc^{(0)}}{1 - i\omega\tau} i \qpl W \left(\frac{\partial \left( -\mu/e \right)}{\partial \Vtg} \right)_{\rhop}.
\end{eq}
(Note that we use $\exp(-i \omega t)$ for the time dependence of harmonic signals instead of $\exp(+j \omega t)$ convention prevalent in electrical engineering, resulting in reactances having unconventional signs.)

The equation on the second order dc current results from the zero dc drain current condition,

\begin{eq}{second-order_channel_current}
\Jloc^{(2)} &= -\frac{1}{e} \sigmadc^{(0)} \frac{\partial \mu^{(2)}}{\partial x} - \frac{1}{2e} \Re\left( \sigmadc^{(1)} \frac{\partial \mu^{(1)*}}{\partial x} \right) = 0,
 \end{eq}
yielding the intrinsic channel responsivity

\begin{eq}{channel_responsivity}
\RC \equiv \frac{\Vout}{\abs{\Vin}^2/2} &\equiv \frac{\left[ \mu^{(2)}(+\infty) - \mu^{(2)}(0) \right]/(-e)}{\abs{\Vin}^2/2}\\
&= - \frac{1}{2} \left(\frac{\partial \ln \sigmadc}{\partial \Vtg}\right)_{\mu} \left(\frac{\partial \left( -\mu/e \right)}{\partial \Vtg} \right)_{\rhop}.
 \end{eq}
 
The expressions \eqref{open-drain_transconductance}, \eqref{channel_responsivity}, and the definition of $\qpl$ \eqref{first-order_channel_solutions} differ from the results of \Cite{supp_Sakowicz} by two extra factors. The first factor $\left(\partial \left( -\mu/e \right) / \partial \Vtg \right)_{\rhop}$ is unity in a single-gated FET and reduces to approximately $\Ct/(\Ct+\Cb) = \db/(\dt+\db)$ in the presence of a bottom gate. The second factor $1/(1 - i\omega\tau)$ appears due to the frequency dependence of conductivity.

Calculations show that the difference between $\left(\partial \left( -\mu/e \right) / \partial \Vtg \right)_{\rhop}$ and  $\db/(\dt+\db)$ is minor and can be neglected within the accuracy of our model, so we used the following expressions for the channel ``source-gate'' conductance and intrinsic channel responsivity:
\begin{eq}{channel_final_results}
\GCSG(\omega) &= - \frac{\sigmadc^{(0)}}{1 - i\omega\tau} i \qpl W \frac{\db}{\dt+\db},\\
\RC &= - \frac{1}{2} \left(\frac{\partial \ln \sigmadc}{\partial \Vtg}\right)_{\mu} \frac{\db}{\dt+\db},\\
\qpl &\equiv \sqrt{\frac{i \omega (1 - i\omega\tau)}{\sigmadc^{(0)}} \frac{\partial \rhop}{\partial \left( -\mu/e \right)}}.
 \end{eq}

The derivatives in \eqref{channel_final_results} were evaluated with the help of the approximate electrostatic model presented in \ref{sec:electrostatics} and the constant-mobility approximation for the channel dc conductivity:
\begin{eq}{channel_dc_conductivity}
\sigmadc^{(0)} = \abs{\rhop} \mu_{\rm BLG} + \sigma_{\rm residual}(\Vbg),
\end{eq}
where we take $\mu_{\rm BLG} = 10^5$~cm$^2$/(V$\cdot$s) (according to measurements performed on similar devices~\cite{supp_our_natcomm}). The transport relaxation time $\tau$ was taken to be 2 ps according to the relation $\mu_{\rm BLG} = e \tau/ m^{*}$, where $m^{*} = \gamma_1/2 v_0^2$ (this is the carrier effective mass in the band extrema of gapless BLG; in gapped BLG band dispersion is similar to the gapless case except in close vicinity of the band edges, so we neglect the bandgap dependence of $m^{*}$).

The residual conductivity $\sigma_{\rm residual}$ due to potential fluctuations in the channel was obtained by fitting the following formula to the experimental dc resistance at the channel neutrality point:
\begin{eq}{residual_conductivity}
\frac{L}{W}\sigma_{\rm residual}^{-1}(\Vbg) = \frac{r_{\infty} \Vbg^2 + r_0 V_0^2}{\Vbg^2 + V_0^2}.
\end{eq}
The fitting procedure yielded $r_0 = 200$~$\Omega$, $r_{\infty} = 150$~k$\Omega$, $V_0 = 5.5$~V.

Using the intrinsic channel responsivity and channel ``source-gate'' conductance, together with the intrinsic tunnel junction and tunnel-gate responsivities and the tunnel junction conductance found in \ref{sec:tunnel_responsivity}, we can obtain the total responsivity of our transistor through \Eqref{total_responsivity} and convert it to the photodetector responsivity through \Eqref{voltage_to_power_responsivity}.

\section{Performance limits of BLG TFET photodetectors}
\label{Limits}
	\begin{figure}[ht!]
	\centering\includegraphics[width=0.33\linewidth]{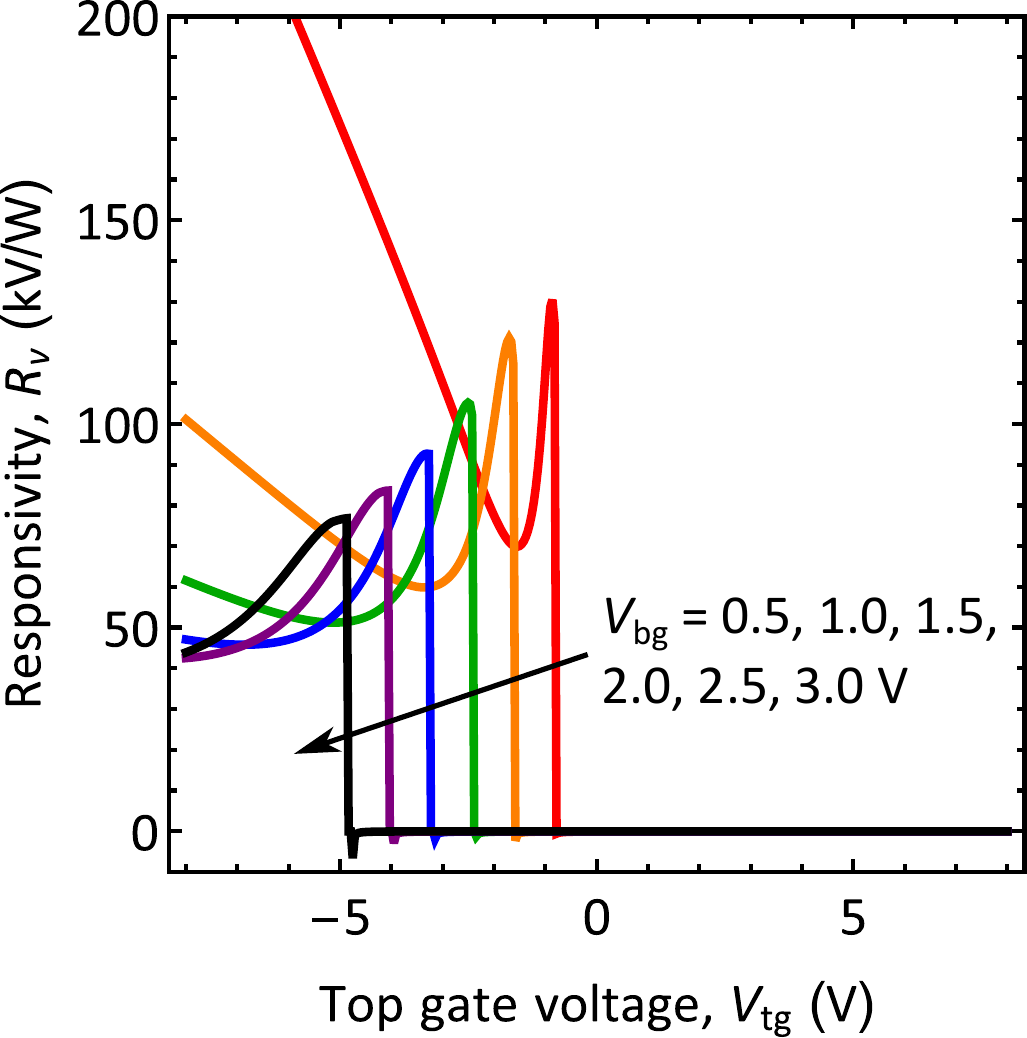}
	\caption{\textbf{Responsivity of an ideal device.} Calculated responsivity of our photodetector in absence of electric potential fluctuations.}
	\label{fig:ideal_responsivity}
    \end{figure}
The theory described in the previous Supplementary Sections was used to calculate the theoretical responsivity of our photodetector, which is shown in Fig.~\ref{Fig4} of the main text. The detector responsivity in our theory is limited by the electric potential fluctuations and could be substantially improved in devices with reduced density of charged impurities. Supplementary Figure~\ref{fig:ideal_responsivity} shows the theoretical responsivity of our photodetector in absence of potential fluctuations (that is, with $\Ffluct =  0$), which reaches hundreds kV/W.
    
Another way to increase detector responsivity is to exploit large nonlinearity of the tunnel junction at small values of band overlap (when the tunnel current is about to be switched off). This requires that the conduction band edge in the source region is simultaneously aligned with the valence band edge in the channel (or vice versa) and with the Fermi level.

Such kind of band alignment can be realized by introducing an additional gate above the source region and could potentially result in a very large nonlinearity even at room temperature~\cite{supp_our_low-voltage_TFET}, which would yield infinite responsivity in the idealized model (no potential fluctuations, no leakage currents). This is easy to show by considering a power-law dependence of the tunnel conductance on the gate voltage, $G_S \propto (\Vtg - V_{\rm th})^{\alpha}$ for $\Vtg > V_{\rm th}$ and zero otherwise, which results from the power-law dispersion $k(E)$ near the band edges. Taking the logarithmic derivative of $G_S$ with respect to the gate voltage, we obtain $\abs{\RSG} = \alpha/(\Vtg - V_{\rm th})$ for nonzero $\alpha$, or a $\delta$-peak at $\Vtg = V_{\rm th}$ for $\alpha = 0$. A similar argument holds for $\abs{\RSS}$, since the band alignment is affected not only by the gate voltage, but also by the Fermi level in the channel.

In practice, the maximum achievable responsivity will be limited by potential fluctuations and leakage currents. Thermionic leakage hinders the performance of our detector at non-cryogenic temperatures because of the small bandgap ($< 60$ meV) realized in our TFET, but this problem can be mitigated by increasing the bandgap, either by applying a larger vertical field to BLG, or by using larger-gap materials, such as black phosphorus. Electric potential fluctiations present a more fundamental issue and limit the logarithmic derivatives of the tunnel conductance to $\sim 1/V_{\rm fluct}$, where $V_{\rm fluct}$ is the magnitude of these fluctuations.

Assuming the total responsivity is dominated by $\Rdiode$ (as in our photodetector) and using equations \eqref{total_responsivity}, \eqref{log_derivative_source}, and \eqref{voltage_to_power_responsivity}, we can estimate the achievable room-temperature noise equivalent power as
\begin{eq}{projected_NEP}
{\rm NEP}_{\rm min} &= \frac{\sqrt{4 r_{\rm 2pt} k_B T}}{\abs{\RVP}} = \frac{\sqrt{4 r_{\rm 2pt} k_B T}}{4 Z_{\rm rad} \abs{\frac{Z_{\rm GS}}{Z_{\rm GS} + Z_{\rm rad}}}^2  \abs{\RVV}} \approx \frac{\sqrt{4 r_S k_B T}}{4 Z_{\rm rad} \left(\frac{r_S}{r_S + Z_{\rm rad}}\right)^2  \frac{1}{2} \abs{\frac{\partial \ln G_S}{\partial V_{sS}}}}\\
&\approx \frac{16}{9}\sqrt{3}\, V_{\rm fluct} \sqrt{\frac{k_B T}{Z_{\rm rad}}},
\end{eq}
where $r_S = G_S^{-1}$ is the resistance of the source tunnel junction. To minimize the noise equivalent power, we assumed $r_S = 3 Z_{\rm rad}$, the drain junction is absent, and the channel resistance is negligible.

Taking $Z_{\rm rad} = 75$~$\Omega$ and $V_{\rm fluct} = 1$~mV (an experimentally achievable value~\cite{supp_fluctuations}), we estimate that the room-temperature noise equivalent power in TFET-based photodetectors can be made as low as 0.02 pW/$\sqrt{\rm Hz}$ (shown in Supplementary Figure~\ref{FigS-comparison}).